\documentclass[fleqn,usenatbib]{mnras}
\usepackage[justification=centering]{caption}
\usepackage{newtxtext,newtxmath}
\usepackage{siunitx}

\usepackage[T1]{fontenc}

\usepackage{breqn}
\DeclareRobustCommand{\VAN}[3]{#2}
\let\VANthebibliography\thebibliography
\def\thebibliography{\DeclareRobustCommand{\VAN}[3]{##3}\VANthebibliography}

\usepackage{graphicx}	
\usepackage{amsmath}	

\title[Model Validation]{Quantifying Interstellar Extinction at High Galactic Latitudes}

\author[M. O'Callaghan et al.]{
Matthew O'Callaghan,$^{1}$\thanks{E-mail:mo503@cam.ac.uk}
Gerry Gilmore,$^{1,2}$
Kaisey S. Mandel$^{1,3,4}$
\\
$^{1}$Institute of Astronomy, University of Cambridge, Madingley Road, Cambridge, CB3 0HA, UK\\
$^{2}$Institute of Astrophysics, FORTH, Crete, Greece\\
$^{3}$Statistical Laboratory, DPMMS, University of Cambridge, Wilberforce Road, Cambridge, CB3 0WB, UK\\
$^{4}$The Alan Turing Institute, Euston Road, London, NW1 2DB, UK}

\date{Accepted XXX. Received YYY; in original form ZZZ}

\pubyear{2024}

\begin{document}
\label{firstpage}
\pagerange{\pageref{firstpage}--\pageref{lastpage}}
\maketitle

\begin{abstract}
{A detailed map of the distribution of dust at high Galactic latitudes is essential for future cosmic microwave background (CMB) polarization experiments because the dust, while diffuse, remains a significant foreground in these regions}. We develop a Bayesian model to identify a region of the Hertzsprung-Russell (HR) diagram suited to constrain the single-star extinction accurately at high Galactic latitudes. Using photometry from Gaia, 2MASS and ALLWISE, and parallax from Gaia, we employ nested sampling to fit the model to the data and analyse the posterior over stellar parameters for both synthetic and real data. Charting low variations in extinction is complex due to systematic errors and degeneracies between extinction and other stellar parameters. The systematic errors can be minimised by restricting our data to a region of the HR diagram where the stellar models are most accurate. Moreover, the degeneracies can be reduced by including {astrophysical} priors and spectroscopic constraints. We show {that} accounting for the measurement error of the data and the assumed inaccuracies of the stellar models are critical in accurately recovering small variations in extinction. {We compare the posterior distribution for individual stars with spectroscopic stellar parameter estimates from LAMOST and Gaia ESO and illustrate the importance of estimating extinction and effective temperature as a joint solution.}

\end{abstract}




\section{Introduction}
Charting the interstellar medium (ISM) provides insight into critical foregrounds for studying Galactic and extra-Galactic astrophysics. As light passes through Galactic dust it is both dimmed and reddened due to scattering and absorption processes collectively known as extinction. The degeneracy between extinction and other stellar parameters implies that constraining stellar parameters is essential to using extinction as a tracer for the structure of the ISM. When extinction is low, as we expect for high Galactic latitude regions, the observed spectral energy distribution (SED) of a star whose starlight has been reddened by the ISM is similar to the observed SED of unreddened starlight emitted by a star with different stellar parameters, such as the effective temperature or metallicity (see {Section} \ref{sed_fit}). In this paper, we highlight the complexities which arise in constraining extinction at high Galactic latitude regions, which we define as regions with Galactic latitude $|b|>45^{\circ}$, when using broadband photometry and parallax observations.

The ISM is a contaminant for future Cosmic Microwave Background (CMB) experiments to investigate the inflationary history of the universe. Current and future cosmological surveys, such as the BICEP and Keck Array \citep{BICEP}, the Simons Observatory \citep{2021AAS...23721403X} and Planck \citep{2020A&A...641A..10P}, aim to accurately constrain cosmological parameters by looking to regions of the sky where foreground contamination is at a minimum, particularly at high Galactic latitudes where interstellar dust is diffuse. In particular, cosmological B-mode experiments are contaminated by the ISM due to starlight {absorbed by asymmetrical dust particles being re-emitted as infrared radiation which is polarised perpendicular to the local magnetic field, given the geometrical properties of Galactic dust \citep{draine_polari}. {Thus, the ISM is a significant foreground for detecting CMB polarization signals and given the sensitivity of the B-mode experiments, a high-resolution map probing the finer structure of the ISM is desirable.}

Dust in the ISM emits preferentially in the infrared. Therefore, the first maps of dust in the ISM, such as the Schlegel-Finkbeiner-Davis (SFD) map \citep{SFD}, derived the column density of dust by charting the thermal emission and provided line-of-sight integrated extinction estimates. The Planck thermal dust map \citep{Planck}, goes further by separating the dust emission from the cosmic infrared background and provides an accurate emission-based integrated dust map. In more recent years, mapping the variation of the extinction of starlight across the sky has been used as a tracer for Galactic dust. The availability of multi-band photometry, spectroscopy and parallax data from Gaia \citep{GAIADR3} has allowed astronomers to create three-dimensional maps of the dust in our Galaxy. {However, constraining extinction using photometry is degenerate with estimating the effective temperature \citep{2011MNRAS.411..435B}, {in particular for low extinction regimes when the effects from this degeneracy have more of a fractional impact on the extinction posterior} (see Section \ref{sed_fit})}. Modern methods have split into two branches: data-driven machine learning methods and Bayesian methods using stellar evolution models. The Bayesian methods (such as \cite{greenmap}, \cite{gspphot}, \cite{starhorse} and \cite{Lallement}) are important to test our understanding of the underlying astrophysical models and incorporate known astrophysics. The machine learning methods usually derive a large solution of data-driven stellar parameters (such as \cite{mlgreen} and \cite{Andrae_2023}) using spectroscopic-derived stellar parameter estimates and extinction estimates from other studies. From here, astronomers can generate dust maps (for example, \cite{mlgreen} and \cite{edenhofer23}), which can provide a far more scalable alternative to Bayesian methods as the amount of data increases. In all photometric dust maps, some underlying stellar model is either assumed or inferred from the data. \cite{gspphot} and \cite{starhorse} use theoretical isochrones for the intrinsic magnitudes/colours and atmospheric models with a mean extinction law to map stellar parameters to data. On the other hand, \cite{greenmap} fits a stellar locus in 7D colour space to a set of stars assumed to have negligible extinction to derive intrinsic colour relations. 

The primary interest of the discussed maps is to provide the large-scale structure of the Galactic dust, mainly toward the Galactic plane where both the density of dust and extinction are high. Many of these dust maps exhibit significant correlation with the large-scale structure of the Galaxy \citep{Mudur_2023}. At high Galactic latitudes, where accurate dust maps on the resolution of future CMB experiments are lacking
\citep{2013ApJ...771...68P}, extinction is low and any signal is sensitively dependent on the systematics and uncertainty when taking measurements. {Generating non-Gaussian dust maps is important for searches of B modes from inflation \citep{2024MNRAS.527.5751A} and a finer structural understanding of Galactic dust is essential for future B mode experiments. The Simons Observatory will map the CMB to 1 arcminute, BICEP operates at 100 GHz and 150 GHz at angular resolutions of $1^\circ$ and $0.7^\circ$ and Planck maps the CMB at resolutions greater than 10 arcminutes. Moreover, the CMB-S4 have a resolution goal of $2$ arcminutes \citep{2022AAS...24021001M}. For each survey's scale of interest, accurate dust distributions along the line-of-sight are important for foreground removal. Moreover, line-of-sight polarisation signals which are extrapolated from higher frequency dust maps to lower, CMB-dominated frequencies may be contaminated by 3-dimensional effects which can decorrelate the polarisation maps between different frequencies, reducing the accuracy of empirical extrapolation models below the level required to detect a primordial B-mode signal \citep{2015MNRAS.451L..90T}. Thus, charting low variation in extinction becomes essential in the context of these surveys.}\\

This is the first paper in a series looking to trace small variations in the ISM by analysing the extinction posterior distribution as it varies across the sky. We focus on determining the full posterior distribution of a single, low-extinction, main sequence star using stellar evolution models. We illustrate the intricate details of the extinction posterior and highlight the degeneracies and systematics that can dominate a low-extinction signal. Analysis of the degeneracies between extinction and other parameters when comparing stellar models to photometric and parallax data is not new (for example, \citealt{2011MNRAS.411..435B}). In this paper, however, the degeneracy analysis is essential to finding a region of the HR diagram where we can accurately constrain extinction. We see this by looking at how the degeneracy depends on what part of the main sequence a star lies on.  Moreover, we include the effect of adding spectroscopic constraints of stellar parameters on the model and analyse the effect of modelling the stellar model inaccuracies when compared to observations. Further papers in preparation will examine applications of using the full extinction posterior at high Galactic latitudes, for example, charting the fine structure of the interstellar medium.

In Section \ref{Data}, we provide an overview of the data used in our model. Section \ref{methods} outlines a method that derives the posterior distribution for extinction. Section \ref{validation_syn} focuses on {the degeneracies within} the model using synthetically generated samples over a large parameter range, showing explicitly {how the degeneracies present themselves} in the posterior. Section \ref{validation_real} validates the model against spectroscopic surveys.

\section{Data Sets}\label{Data}
In this section, we describe the surveys used in this paper. The data products from Gaia, 2MASS, ALLWISE and spectroscopic surveys will be used directly in our calculation of extinction posteriors.
\subsection{Gaia DR3}
We use position, parallax and photometric magnitudes made available by the European Space Agency's (ESA) Gaia mission \citep{GAIA_MISSION}. Gaia is a pioneering survey that aims to provide a precise three-dimensional chart of more than a billion stars throughout our Galaxy, mapping their distances, motions, luminosity, composition, and temperatures. The satellite consists of a broad $G$ band filter (330-1050 nm), a $BP$ (330-680 nm) and $RP$ (630-1050 nm) low-resolution fused-silica prisms and a Radial Velocity Spectrometer (RVS) instrument. In 2022 the Gaia Data Processing Consortium released the Gaia Data Release 3 \citep{GAIADR3}, which provides a full astrometric solution (position, parallax and proper motions) for around 1.46 billion sources, together with photometric apparent magnitudes $G$, $G_{BP}$ and $G_{RP}$. We use all apparent magnitudes from Gaia and parallaxes as observations in our model. We note that the Gaia BP and RP absolute magnitudes are derived from integrating the Gaia BP and RP spectra, respectively. We choose not to use the Gaia BP/RP spectra themselves due to the difficulty in accounting for systematic error in mapping a synthetic SED to the BP/RP spectra and modelling this process is an active research area. We apply the parallax zero-point corrections as recommended in \cite{zero-point}, which depend on position, magnitude, and colour. Next, as recommended by \cite{phot-validation-gaiaedr3}, we select sources which have apparent magnitudes $m_{BP}<20.5$ mag or $m_{RP}<20.0$ mag. Furthermore, we make a cut using the renormalized unit weight error (RUWE) and set it to be $<1.2$. 

{The large widths of the Gaia photometric passbands mean that} their extinction coefficients, $A_X/A_0$, are temperature dependent. It can be useful to include narrower passbands in the optical wavelength range to further constrain the extinction posterior. We mention the effect of including the Pan-STARRS \citep{PS1} photometric data in our model in Appendix \ref{panstarrs}.

\subsection{2MASS and ALLWISE}
We use the Gaia cross-matching recommendations \citep{CROSSMATCH} to cross-reference Gaia with both the 2MASS and ALLWISE surveys. The Two Micron All Sky Survey (2MASS; \cite{2MASS}) is a ground-based, all-sky survey in three near-infrared passbands,  J (1.25 $\mu $ m), H (1.65 $\mu$ m), and Ks (2.16 $\mu $ m) with a $1\sigma$ photometric uncertainty of <0.03 mag \citep{2mass06}. In this paper, we use all of the 2MASS passbands and we only select cross-matched sources with the highest 2MASS photometric ('AAA') quality flag. The Wide-field Infrared Survey Explorer (WISE; \cite{WISE}) mapped the sky in the W1 (3.4 $\mu m$), W2 (4.6 $\mu m$), W3 (12 $\mu m$), and W4 (22 $\mu m$) bands. The AllWISE Data Release \citep{ALLWISE}, includes all of the photometry from the two WISE phases and NEO-WISE \citep{neowise}. The AllWISE Source Catalog consists of photometry of over 747 million objects. We only take the W1 and W2 passband photometry in this paper and, similarly to 2MASS, we select cross-matched sources with the highest photometric ('AA') quality filter.

\subsection{Spectroscopic Surveys}

In this paper, we use spectroscopic-derived stellar parameters {(effective temperature, metallicity and log surface gravity)} as benchmarks to validate our model. High-resolution and low-resolution spectroscopic surveys, along with their associated data products can be used explicitly as constraints in our model to help constrain the extinction posterior. In this paper, we employ high-resolution spectroscopic features to establish a best-case scenario and compare the extinction posterior with that obtained from our model.

We use the highly accurate Gaia-ESO iDR6 \citep{2022ges} as a benchmark survey to validate our model in Section \ref{validation_real}.  {Gaia-ESO is particularly noteworthy for its extinction-independent \citep{2015A&A...577A..77S} estimates of the stellar parameters and their uncertainty.} {We also} use LAMOST DR8 \citep{LAMOSTDR8} derived parameters to define a prior on the metallicity parameter in Section \ref{methods} and to validate our model in Section \ref{validation_real}.

Moreover, we will compare the extinction posterior to stellar parameters obtained when using constraints derived from the Gaia BP/RP (XP) spectra, in Section \ref{validation_real}.

\section{Methods}\label{methods}
Models of stellar evolution demonstrate that a star's first-order measurable properties are determined by its initial mass ${m}$, initial metallicity $[Fe/H]$, and age $a$, particularly for stars evolving along the main sequence of the Hertzsprung-Russell (HR) diagram where our understanding of stellar evolution is the most robust. In this paper, the initial mass and metallicity are given as multiples of the solar values, and age is defined in years. Given an initial value of mass and metallicity, stellar evolution codes evolve a stellar model until it reaches a desired age, predicting other stellar parameters such as the effective temperature ($T_{\text{eff}}$), radius ($R$) or the log surface gravity $\log(g)$, for example. We define our model parameters as $d$, $A_0$, $ R_V $, and $\boldsymbol\theta = ({m},[Fe/H],\log(a))$, where $d$ is the radial distance in parsecs, $A_0$ is the monochromatic extinction at $541.4$ nm, (approximately the centre of the $V$ band) and $R_V=A_V/(A_B-A_V)$ is the total-to-selective extinction parameter. 

Throughout, we will refer to $A_0$ as the extinction (or the extinction parameter) and $R_V$ will be referred to as the total-to-selective extinction. We propose a forward model which, given a sample of $\boldsymbol\theta$, $d$, $R_V$ and $A_0$, predicts the apparent magnitudes (from the Gaia, 2MASS and ALLWISE passbands) and parallax. Moreover, we allow for the use of spectroscopic constraints of stellar parameters by assuming a Gaussian error on the spectroscopic constraint. By comparing this against a dataset of noisy photometric and parallax measurements we infer the stellar parameters by computing the posterior distribution on a star-by-star case.

\subsection{Stellar Model}
Our forward stellar model is a function  $(\boldsymbol{{x}}, \boldsymbol\Theta)=M(\boldsymbol\theta, d, A_0, R_V )$ from the model parameters to apparent magnitudes $\boldsymbol{{x}}$ and stellar parameters $\boldsymbol\Theta$. If $\boldsymbol\Theta$ contains any of the stellar parameters in $\boldsymbol\theta$ then the mapping is just the identity between those components. The forward model consists of two main components: a model of stellar evolution provided by a grid of evolution tracks (or isochrones) and synthetic templates of stellar spectra. {\cite{2011MNRAS.411..435B} shows that using an astrophysical HR diagram prior is useful in generating absolute magnitudes and reconstructing stellar parameters. The stellar evolution model} is included by performing a 3D interpolation along a grid of stellar parameters provided by a collection of isochrones to get a consistent estimate of the derived stellar parameters $(T_{\text{eff}},\log R, \log(g))$ and luminosity given a value of $\boldsymbol\theta$. We use the Mesa Isochrones and Stellar Tracks (MIST; \cite{Choi_2016}) to define our stellar model grid. The MIST website provides pre-packaged model grids and outlines an accurate method of interpolating to new parameters and generating isochrones.

To first-order the spectrum of a star $F(\lambda)$ is determined by the effective temperature, metallicity and surface gravity. For the stellar spectra templates grid, we use the ATLAS9 synthetic atmosphere models \citep{2003IAUS.210P.A20C} and perform a cubic interpolation over the grid which, for given stellar parameters, provide intrinsic flux values $F(\lambda)$ for a range of wavelengths. The atmosphere models have an effective temperature minimum of $T_{\text{eff}}>3500K$, but MIST provide stellar parameters and a bolometric correction grid for stars with a lower effective temperature. While we use this extrapolation for illustrative purposes in our inference we discard all stars which return a $T_{\text{eff}}$ value under this bound. We model the effect of extinction by using the \cite{f99} parameterisation of the wavelength-dependent extinction law, which causes the flux to transform as $f(\lambda)=F(\lambda)10^{-0.4 A(\lambda,A_0,R_V)}$. We tested our model also using the PHOENIX \citep{phoenix} atmosphere models and found high agreement in photometry calculations, with small systematic offsets in the stellar parameters of best-fit. In Section \ref{evolution_model} we discuss the systematics that arise from these choices and show that for low-extinction, main sequence stars the effect on the extinction distribution is negligible.

Given a sample of the model parameters, the forward model runs by first interpolating over the stellar evolution tracks to compute consistent estimates of $(T_{\text{eff}},\log R, \log (g))$. With these values, we interpolate over the stellar atmosphere grid and apply the extinction law to derive the extinguished flux $f(\lambda)$. Using the value of the radius and line-of-sight distance we scale the flux to convert it to the flux we would observe by scaling by $(R^2/d^2)$. Then we use the transmission functions $T(\lambda)$ for Gaia \citep{gaia_passband}, 2MASS \citep{2mass_passband} and ALLWISE \citep{Kirkpatrick_2016} to derive the apparent magnitude for a source as \begin{equation}
    x=-2.5\log_{10}\left(\frac{\int \lambda T(\lambda)F(\lambda)(R^2/d^2)10^{-0.4 A(\lambda,A_0,R_V)}\text{d}\lambda }{\int \lambda T(\lambda)\text{d}\lambda }\right) + ZP,
\end{equation}
where $ZP$ are the Vega zero points associated with the passband.

\subsection{Likelihoods}
Given noisy observations of the apparent magnitudes ($\boldsymbol {\hat{x}} $), parallax ($\hat \omega$), {on-sky Galactic longitude ($\hat{l}$) and latitude ($\hat{b}$)},  and, optionally, spectroscopic constraints of a subset of $p$ stellar parameters centred on $\hat{\boldsymbol \Theta}=(\hat{\Theta}_1,...,\hat{\Theta}_p)$,  (effective temperature or metallicity, for example), we wish to compute the posterior distribution of the extinction parameter, $A_0$, for a single star: 
\begin{equation}
    P(A_0|\,\boldsymbol {\hat{x}},\hat{\boldsymbol \Theta},\hat\omega,\hat l,\hat b)=\int \text{d} d \int \text{d}R_V \int \text{d}\boldsymbol\theta \,P(A_0, \boldsymbol \theta , d, R_V |\, \boldsymbol {\hat{x}},\hat{\boldsymbol \Theta},\hat\omega,\hat l,\hat b).
\end{equation}
Using Bayes' Theorem we have
\begin{equation}\label{post}
\begin{aligned}
P(A_0, \boldsymbol \theta , d, R_V |\, \boldsymbol{\hat{x}},\hat{\boldsymbol \Theta}, \hat\omega ,\hat l,\hat b) &= \frac{1}{P(\hat{x},\hat{\boldsymbol \Theta}, \hat\omega |\,\hat l,\hat b)}P(\boldsymbol{\hat{x}}|\,A_0, \boldsymbol \theta , d, R_V )\\
&\times P(\hat{\boldsymbol \Theta}|\,\boldsymbol\theta) P(\hat \omega |\, d ) P(A_0, \boldsymbol \theta , d, R_V |\, \hat l,\hat b).
\end{aligned}
\end{equation}
We let $(\boldsymbol{x}, \boldsymbol{\Theta})=M(A_0, \boldsymbol \theta , d, R_V )$ be the output of the forward model (note that $\boldsymbol\Theta$ only depends on $\boldsymbol\theta$),  {and assume the joint likelihood of $(A_0, \boldsymbol \theta , d, R_V )$ and the likelihood of $d$ are given by the} {Gaussian normal distributions}:\\
$P(\boldsymbol{\hat{x}}|\,A_0, \boldsymbol \theta , d, R_V )=N(\boldsymbol{\hat{x}} |\, \boldsymbol{x}, \boldsymbol{\Sigma}),$
\begin{equation}P(\hat \omega|\, d)=N(\hat \omega |\,1/d - \omega_{zp},\sigma_\omega^2),\end{equation} 
{where $\omega_{zp}$  is the Gaia-provided parallax zero-point offset.}
We {can use spectroscopic constraints of stellar parameters in the likelihood} by assuming the estimated values are sampled around the truth with some Gaussian error: $
    P(\hat{\boldsymbol \Theta}|\,\boldsymbol\theta)=N(\hat{\boldsymbol \Theta}|\,{\boldsymbol \Theta},\text{diag}(\sigma_{\Theta_1}^2,...,\sigma_{\Theta_p}^2)).
$
We will define our default model as one which has no spectroscopic constraints.

Gaia provides a Python routine to calculate the parallax zero point $\omega_{zp}$ for each source \citep{Lindegren_2021}. {We also define an error floor by setting $\boldsymbol{\Sigma}_{ii}= \sigma_{\text{floor}}^2+\sigma_i^2$,} {where $\sigma_{\text{floor}}=0.05$ mag by default, is} in the photometric magnitudes to account for any error in the interpolation, passband definition, and stellar model systematics. We choose this value from analysis in Section \ref{evolution_model}. However, we stress that this value directly affects the width of the extinction posterior. A smaller value is desirable, but a value which underestimates the systematics will produce an incorrect extinction distribution. Further calibration must be carried out if we wish to constrain this value. We illustrate the effect {of $\sigma_{\text{floor}}$} on the posterior in Section \ref{validation_syn}.

\subsection{Astrophysically Informed Priors}\label{a_informed_p}

{We {adopt} a prior distribution $P(A_0, \boldsymbol \theta , d, R_V |\, \hat l,\hat b)$ as a function of the on-sky coordinates, which we assume are known exactly. Moreover, we assume conditional independence of the model parameters such that the posterior distribution factorises as }\begin{equation}
\begin{aligned}    
    P(A_0, \boldsymbol \theta , d, R_V |\, \hat l,\hat b)&=P(A_0)P(R_V)P(m)P(\log({a})|\, \hat l,\hat b)\\
    &\times P([Fe/H] |\, d, \hat l,\hat b)P(d |\, \hat l,\hat b)
\end{aligned}
\end{equation}
{Throughout this paper we compare the effects of using uninformed priors and astrophysical priors, which we define as distributions drawing from either theoretical models or astronomical data. In particular,} we adopt {astrophysical priors} relevant to the high Galactic latitude regions (which depends on $\hat{l}, \hat{b} $).{ From \cite{Bailer_Jones_distance} we adopt the Generalised Gamma distance prior,

\begin{equation}\label{bjp}
P(d|\, \hat l,\hat b)=\begin{cases}
        \frac{\alpha}{\Gamma(\alpha^{-1}(\beta+1))} \frac{d^\beta}{L^{\beta+1}} \exp(-(\frac{d}{L})^{\alpha}) & \text{if } d >0 \, \rm pc\\
        0 & \text{if } d=0  \, \rm pc
    \end{cases}
\end{equation}
{where the values of $\alpha, \beta$ and $L$ depend on the sky position, $\hat l, \hat b$, and were {fit} to a three-dimensional model of our Galaxy. These values can be }queried from the accompanying data products of \cite{Bailer_Jones_distance}. For the prior on the initial mass, we adopt the \cite{2003PASP..115..763C} initial mass function (IMF). 

The prior distribution of metallicity is modelled as $p([Fe/H] |\, d, \hat l,\hat b)$ from the LAMOST \citep{LAMOSTDR8} $[Fe/H]$ values for sources at high Galactic latitudes which have valid distance estimates from Gaia, using a Kernel Density Estimate. {We display a distance-binned plot of the LAMOST metallicities at high Galactic latitudes in Figure \ref{fig:feh_prior_data}.}

We focus on regions of high Galactic latitude, at apparent magnitudes where the star counts are dominated by old main sequence stars and the extinction is expected to be low. Therefore, the $\log(a)$ parameter is modelled with a uniform prior with bounds $(8.9,10.4)$ $\log(\rm yrs)$. 

The extinction parameter, $A_0$, is modelled with an exponential prior with exponential scale parameter {$d_A$. A}t high Galactic latitudes, we expect extinction to be small {but, for a given star, the prior on extinction can be informed by the Planck \citep{Planck} integrated emission map for that region. If we use this information then we choose $d_A$ so that there is low prior probability for values of $A_0$ exceeding this value. This allows for extinctions greater than the Planck value as we wish to find variation in extinction over a finer region than the Planck map has integrated over.} Based on analysis from \cite{zhang2023rv}, the $R_V$ prior is a normal distribution with a mean of 3.1 and a standard deviation of 0.25 to account for any variation in the extinction law. For an overview of the priors see Table \ref{tab:priors}.

\begin{figure}
    \centering
    \includegraphics[width=0.5\textwidth]{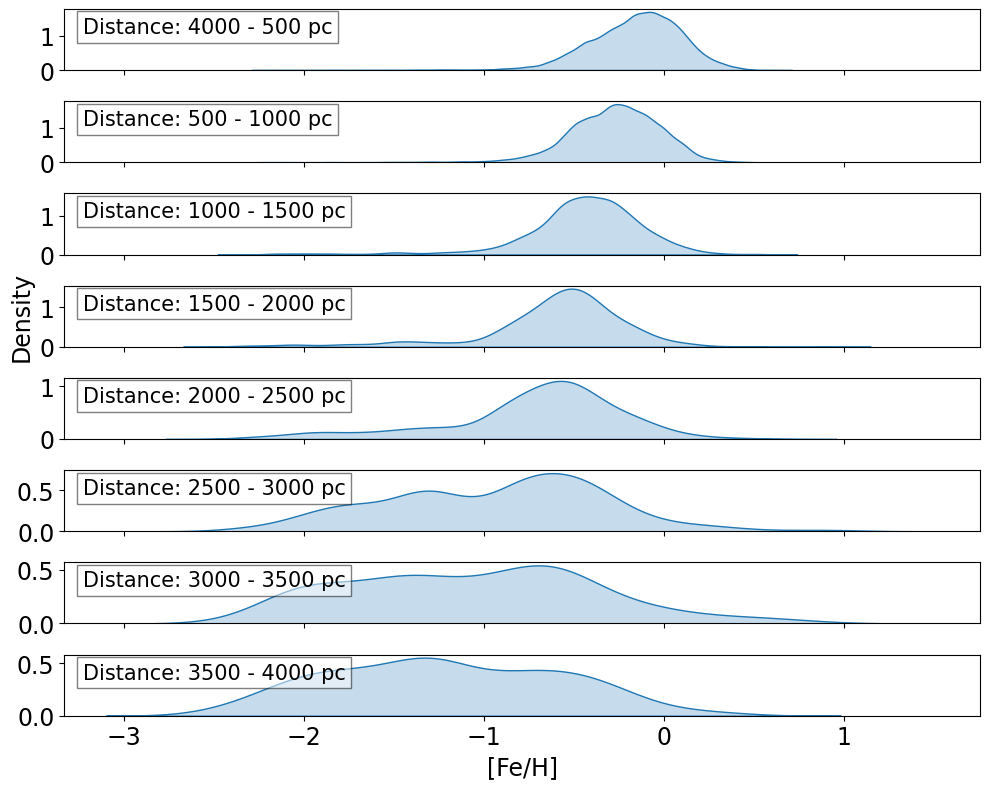}
    \caption{Distribution of LAMOST metallicities {(in dex)} at $|b|>45^\circ$ as a function of different distance bins. We use a KDE to model the metallicity distribution conditional on distance.}
    \label{fig:feh_prior_data}
\end{figure}

\begin{table}
  \centering
  \begin{tabular}{||c c c||}
  \hline
    \textbf{Parameter} & \textbf{Prior Distribution} & \textbf{Description} \\
    \hline
    \hline
     $\log(a)|\, \hat{l}, \hat{b}$& U($8.9$, $10.4$) {$\, \rm \log(yr)$}& Logarithm of stellar age \\
     \hline
    $m$ & \cite{2003PASP..115..763C} IMF & Initial Mass \\
    \hline
    $[Fe/H] |\, d, \hat{l}, \hat{b}$& Modelled on LAMOST & Metallicity \\
    \hline
    $d|\, \hat{l}, \hat{b}$& \cite{Bailer_Jones_distance}  & Stellar distance \\
    \hline
    $A_0$ & exp$(1/d_A)$& Extinction \\
    \hline
    $R_V$ & $N(3.1, (0.25)^2)$ & Extinction Ratio \\
    \hline
    \hline
  \end{tabular}
  \caption{Prior Distribution for Model Parameters.}
  \label{tab:priors}
\end{table}

\subsection{Uniform Priors}\label{uniform_p_sec}
{{In Section \ref{validation_syn}, we validate our model using different uniform priors. In particular, we will consider a uniform prior on the metallicity $[Fe/H]\sim U(-3,0.5)\, \rm dex$ and extinction $A_0 \sim U(0,4)\, \rm mag$ to investigate the consequences of using uniformed priors and how sensitive the data are to extinction.} In Section \ref{uniform_priors}, \ref{floor_error}, \ref{teff_constraint} and \ref{transform} we assume the uniform prior distribution on both the metallicity and extinction. In Section \ref{metallicity_prior} we assume the uniform extinction prior. } Elsewhere, we use the priors as outlined in Section \ref{a_informed_p}.

\subsection{Nested Sampling Procedure}
Nested sampling \citep{nestedsampling} is a Monte Carlo method to estimate the evidence ($P(\hat{\boldsymbol{x}},\hat{\boldsymbol \Theta}, \hat\omega)$ in Equation \ref{post}) and generate samples from the posterior distribution by transforming the evidence integral into a one-dimensional integral and sampling 'live points' to derive an estimate. Nested sampling can sample from the tails of the posterior in a reasonable run-time and is capable of drawing samples from multi-modal posteriors with degenerate parameters \citep{Buchner_2023}, and for a fixed architecture outperforms the traditional MCMC algorithms, like Metropolis-Hastings \citep{1970Bimka..57...97H}, in sampling from multi-modal posteriors. More advanced methods have been developed since its inception, including MultiNest \citep{Feroz_2009}, PolyChord \citep{2015MNRAS.450L..61H}, dynesty \citep{2020MNRAS.493.3132S} and UltraNest \citep{buchner2021ultranest}.
Nested sampling is ideal for our problem, where the posterior is multi-modal due to the model parameters being highly degenerate with each other. We use the UltraNest Python package \citep{buchner2021ultranest} to perform a nested sampling procedure and sample from our Bayesian model. We note that not all combinations of the prior parameters provide valid solutions to the stellar evolution code and in this case, we set the log-likelihood to be $l=-1\times 10^{11}$. We can do this because nested sampling iteratively removes points with the least likelihood.

\subsection{Discussion of Some Systematic Effects of Varying Our Model}\label{evolution_model}

\subsubsection{Changing the stellar evolution model}
When incorporating a stellar evolution model into our Bayesian model there is an assumption about the underlying physics made. We choose the MIST isochrones, with a rotation parameter of $v/v_{crit}=0.4$, and the \cite{f99} extinction law. We note that for old main sequence stars the difference between rotating and non-rotating models is negligible (see Figure 8 in \cite{MIST}).

Other isochrones and stellar evolution tracks exist, such as the PARSEC \citep{parsec} and BaSTI \citep{basti} evolution models and isochrones. Key differences arise in pre-main sequence and post-main sequence phases. For example, The MIST isochrones include the thermally pulsing asymptotic giant branch (TP-AGB) as a phase of the evolutionary code \citep{Choi_2016} and the C-star sections of the MIST isochrones start at a later point along the TP-AGB than for the PARSEC isochrones \citep{Marigo_2017}. Comparing the MIST, BaSTI and PARSEC isochrones shows excellent agreement near and on the main sequence especially at lower metallicities \citep{basti}. However, at higher metallicity, the lower masses which, at younger ages, are still evolving along the pre-main sequence phase are significantly systematically different. We see this effect when we compare a large range of MIST-assigned main sequence synthetic stars from PARSEC and MIST in Figure \ref{fig:compareisochrones}. We see that at older ages, for a given mass, the combination of parameters that MIST has allocated as a main sequence star has been evolved by PARSEC. However, if we choose to select a main sequence star from the PARSEC isochrones with a fixed age and mass, we can generate a dense MIST isochrone of the same age and metallicity {and find a star along the main sequence with different mass where the predicted photometry is almost identical to the PARSEC model. We also see a low mass discrepancy between the different sets of isochrones. For example, see the solar isochrones from MIST and PARSEC displayed in Figure \ref{fig:solarisochrones}. This arises mainly due to the use of \cite{2003IAUS.210P.A20C} models, which are known to be unreliable temperatures lower than $4000 \, \rm K$ \citep{Vines_2022}.}

\subsubsection{Changing the assumed extinction law}

Our choice of the \cite{f99} extinction curve could be replaced with the \cite{ccm} (or any other) extinction curve. In comparing these two curves, a difference {(which depends on the region of interest of the HR diagram)} arises which impacts the computation of the Gaia photometry, with a {maximum difference in the extinction coefficient using the two different extinction laws, $\Delta A_G/A_0$, of around 0.04 when $A_0=1 \, \rm mag$}. In our study, this, while worthy of note, does not cause significant issues. Our focus on low extinction regions of the ISM minimises this systematic where we wish to look for relative differences in the extinction. Moreover, in the other passband, this effect is even smaller. {Any of the above systematics can be modelled by changing $\sigma_{\text{floor}}$ to account for the propagation of the systematic to the predicted photometry.}

 \begin{figure}
    \centering
    \includegraphics[width=0.5\textwidth]{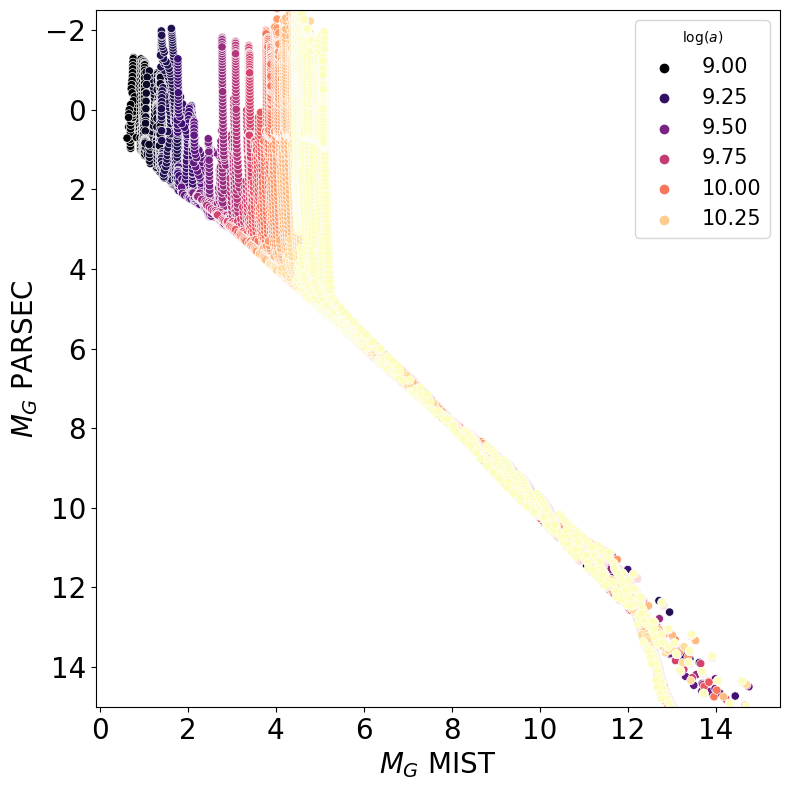}
    \caption{{Absolute Gaia $G$ magnitude from PARSEC vs MIST for a range of mass, metallicities and log-ages, with $\log(a)$ hue in $\log(\rm yrs)$. We match the output of PARSEC and MIST models for each input set of parameters and display the points that MIST allocates as main sequence stars. We do not cut the PARSEC isochrones, to investigate where they differ. We see at brighter absolute magnitudes that PARSEC allocates the stars as post-main sequence stars. For a fixed age and metallicity, PARSEC will allocate a star to be a post-main sequence star at a lower mass than MIST.}}
    \label{fig:compareisochrones}
\end{figure}

\begin{figure}
    \centering
    \includegraphics[width=0.5\textwidth]{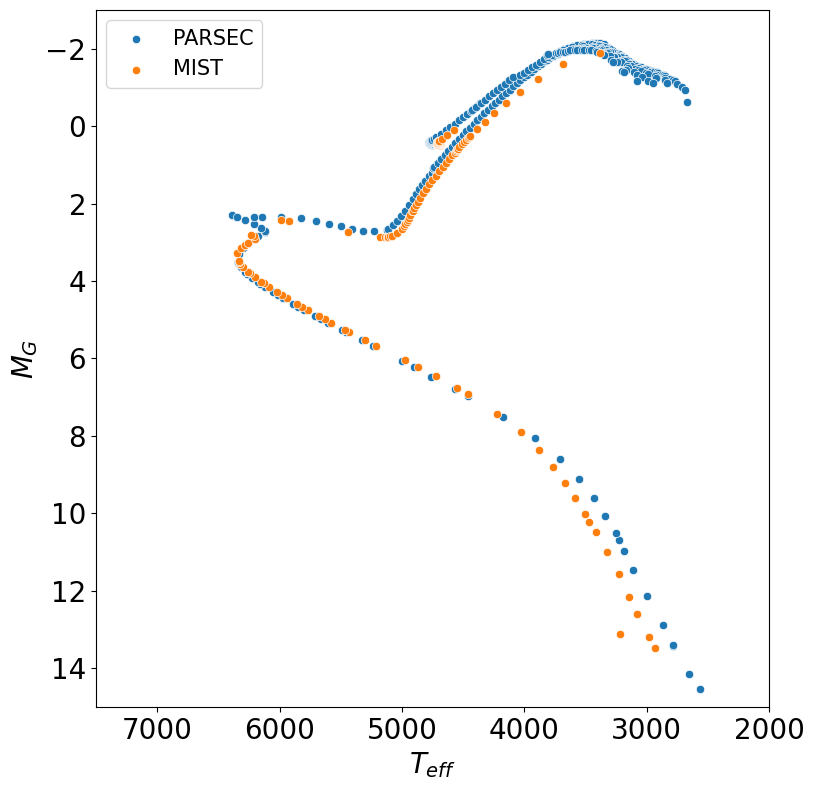}
    \caption{{Absolute Gaia $G$ magnitude versus effective temperature (K) for an isochrone with solar metallicity and age generated using MIST and PARSEC.} We see that they {agree well along the main sequence for $T_{\rm eff}>4000 \, \rm K$.} At brighter absolute magnitudes, PARSEC assumes certain masses have evolved to post-main sequence stars while MIST assumes they are still main sequence stars. We will make cuts to focus on regions where the different stellar evolution models agree.}
    \label{fig:solarisochrones}
\end{figure}

\subsubsection{Changing the prior on stellar ages}
Our model is designed for the analysis of high Galactic latitude stars, which are mainly old. With increasing age, the evolution of sources to a post-main sequence phase happens at lower absolute magnitudes. In Figure \ref{fig:compareisochrones}, we can see in the upper limit our models begin evolving to a post-main sequence phase at around $M_G=4 \, \rm mag$. At magnitudes fainter than this, the age parameter does not change the derived photometry to a large degree. It is important to note, however, that for stars nearer to the solar neighbourhood we expect to find stars younger than those at further distances. However, in this region we expect extinction to be very low. Moreover, for the absolute magnitude cuts we will define in the next section, we test our model on finding the extinction distribution for stars younger than our prior and the systematics are negligible.

\section{Degeneracy Analysis}\label{validation_syn}
In this section, we validate our model on synthetic data sets by using the MIST isochrones \citep{MIST} to generate synthetic photometry and noisy constraints of the stellar parameters, with the constraint widths akin to current Gaia XP spectra parameter retrieval (in particular \cite{Andrae_2023}). We begin by recalling that we are focusing on high Galactic latitudes, where the distribution of stars is dominated by old main sequence stars and our synthetic populations reflect this.

\subsection{Reliable Population of High Galactic Latitude Stars for Constraining Extinction}
{We analyse the stellar evolution models used in our forward model to identify a region of the HR diagram that is occupied by old main-sequence stars for which the stellar models are robust and reliable.}

We restrict our stellar models to the region of the HR diagram that is consistent with the allowed parameter ranges from the \cite{2003IAUS.210P.A20C} stellar atmosphere models so we can interpolate over them in [Fe/H] - $\log g$ - $T_{\text{eff}}$ space. We compare this range with the parameters generated by the MIST isochrones for an age range between $10^{8.9}$ and $10^{10.3}$ years. The range of $\log g$ provided by the stellar atmosphere models is from 0.0 to 5.0 dex and the effective temperatures have a lower bound of $3500$ K. To ensure that our interpolation scheme provides a valid solution {and does not extrapolate outside the reliable parameter range} we restrict ourselves to $3.3<\log g <4.9$ and $3750 \, \rm{K}<T_{\text{eff}}$, where the lower bound on $\log g$ is to mitigate against post-main sequence stars.

\begin{figure*}
    \centering
    \includegraphics[width=\textwidth]{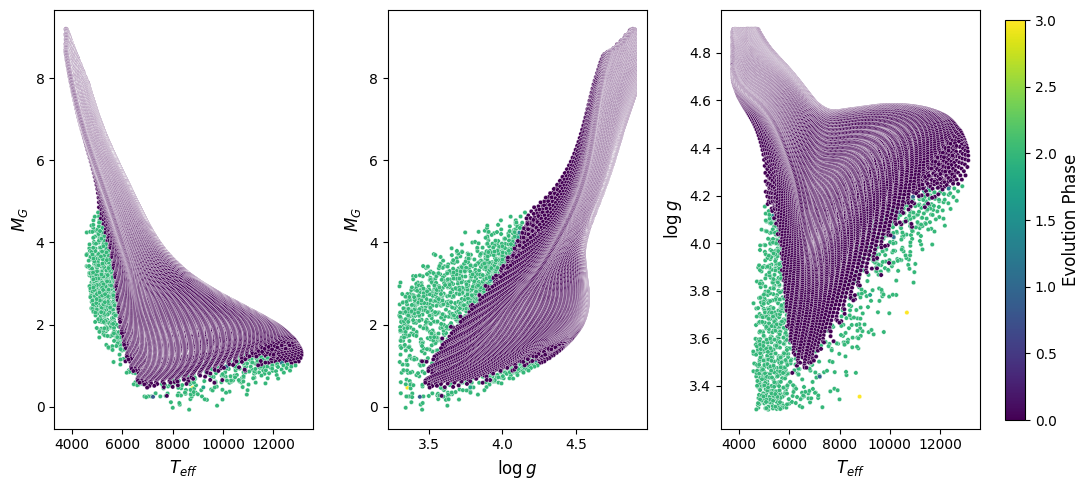}
    \caption{Gaia absolute magnitude $M_G$ versus the MIST effective temperature {(K) and log-surface gravity (dex)}. We illustrate the credible parameter range from stellar atmosphere models, showing where we expect to find post-main sequence stars in the HR diagram. The hue indicates the phase of evolution in the HR Diagram where 0 corresponds to the main sequence.}
    \label{fig:par_range}
\end{figure*}

In Figure \ref{fig:par_range}, we show the result of making the above parameter cuts to the MIST isochrones with $8.9 \leq \log(a) \leq 10.3 \, \log(\rm yrs)$, and illustrate the region of the Gaia HR diagram where valid solutions are feasible. We also make an initial mass cut, $m>0.4 \, M_\odot$, to reduce the chance of {selecting} fully convective stars. We wish to restrict the region further to minimise the {selection} of post-main sequence stars. This amounts to avoiding regions of the HR diagram in Figure \ref{fig:par_range} where we find sources with a phase greater than $0$. Making a corresponding cut to observed absolute magnitudes will increase the probability of {select} main sequence stars. This has two main benefits; firstly, we are restricted to a region of the HR diagram where stellar evolution theory is most robust and the least systematics are present in choosing the model (see Section \ref{evolution_model}). Secondly, it is in this region of the HR diagram that the majority of the stars in the thick disk (and a significant proportion in the thin disk) at high Galactic latitudes appear. From looking at how the stellar parameter cuts propagate to Gaia absolute $G$ magnitude $(M_G)$ space, we can define a region of the HR diagram where the stellar models are the most robust for use in our forward model. We see that these cuts correspond to selecting sources with approximately $4.0<M_G<8 \, \rm mag$. 

Theoretically, our model can deal with determining the extinction of post-main sequence stars {but stellar models are less accurate at reproducing the observables of post-main sequence stars.} When fitting to real data the contribution of an increased error in theoretical modelling the later stages of a star's life is beyond the scope of this work and not scientifically relevant to our primary focus. Our model would provide an underestimation of the error on $A_0$ in this region, which could be a large contaminant when we are looking for small variations in extinction.

If we have a reliable constraint on the surface gravity across the dataset we can make the cut more specific to allow for brighter absolute magnitudes. However, the discussed cut in the Gaia $G$ absolute magnitude should mitigate against finding post-main sequence and fully convective stars. When running the nested sampling procedure, an important flag will be raised if a significant proportion of the posterior mass for the effective temperature and log surface gravity is near the boundaries of the cuts described in this section.

\subsection{Degeneracy Analysis from Fitting Spectra}\label{sed_fit}
{We illustrate the degeneracies which arise in fitting medium-resolution stellar spectra and their influence in low-extinction regimes. In subsequent sections, we proceed to understand how these degeneracies present themselves using our photometric model. {The degeneracy is analysed through a likelihood analysis and does not include any prior distributions.}}

We use the ATLAS9 \citep{2003IAUS.210P.A20C} and PHOENIX \citep{phoenix} grid of synthetic spectra resampled to a spectral resolving power of $\lambda/\Delta \lambda = 720$, and the \citep{f99} extinction law with $R_V=3.1$ to generate synthetic fluxes at different wavelengths over a grid of stellar parameters. We define our grid by setting $\log (g)=4.5$ dex (we found similar conclusions for different $\log (g)$ values) and allowing the effective temperature, metallicity and extinction range between $3500 <T_{\text{eff}}<8000 \,\text{ K}$, $-2.5<[Fe/H]<0.5 \,\text{ dex}$, and $0<A_0<2 \, \rm mag$, respectively. For values not defined by the original grid we perform cubic interpolation over grid points to generate fluxes at the given wavelengths for the corresponding grid point. This defines a vector function $\textbf{f}=f([Fe/H],T_{\rm eff},\log(g),A_0)$, which outputs fluxes at a set of fixed wavelength points. We set a lower bound of the wavelengths to be the Lyman limit, $912$ {\AA} and an upper bound of $65,000$ {\AA} (larger than the upper bound of the W1 and W2 passband filters). Throughout, we will resample the spectra to several wavelength observations to investigate how different observations of the SED affect the accuracy we can recover extinction. 

We illustrate the degeneracies which arise in fitting the synthetic spectra, $\textbf{f}=f([Fe/H],T_{\rm eff},\log(g),A_0)$, to an observed spectrum, $\hat{\textbf{f}}$, assuming a Gaussian likelihood function with error covariance matrix $\boldsymbol{\Sigma}$ arising from the measurement error of the flux at each wavelength. We can find estimates of the parameter covariances for the maximum likelihood estimate using the Fisher information matrix. Throughout, we assume the measurement error of the flux at each point constitutes the Gaussian covariance matrix $\boldsymbol{\Sigma}=\hat{\textbf{F}}\times(2\log_{e}(10)(0.01)/5)^2 $, where $\textbf{F}$ is the diagonal matrix with $\hat{F}_{ii}=\hat{f}_i^2$. This flux error corresponds to a constant photometric magnitude error of $0.01$ mag. Fixing all parameters except two, $\alpha$ and $\beta$, the Fisher information matrix $I$ is given by the components $I_{\alpha,\beta}=\frac{\partial \textbf{f}}{\partial \alpha}^T\boldsymbol{\Sigma}^{-1}\frac{\partial \textbf{f}}{\partial \beta}$. If we define $\boldsymbol{C}=I^{-1}$, we introduce covariances for the parameters which, for the maximum likelihood estimate, satisfy the Cramer Rao lower bound. Thus, we define $\sigma(\alpha)=\sqrt{C_{\alpha,\alpha}}$ and $\rho(\alpha,\beta)=C_{\alpha,\beta}/{\sigma(\alpha)\sigma(\beta)}$, as the standard deviation of the parameter $\alpha$ and the correlation between parameters, respectively.

There exists a degeneracy between the effective temperature and the extinction parameter across the whole grid of stellar parameters. For a fixed measurement error, observing fewer fluxes reduces the ability to constrain the effective temperature and extinction. Thus, increasing the number of passband filters and independent observations in ones analysis will provide a better estimate of the extinction. {To illustrate this, 
 we fix $[Fe/H]$ and $\log (g)$ dex to inspect the relationship between the effective temperature and the extinction parameter for a range of stars. We evaluate $\textbf{f}=f([Fe/H],T_{\rm eff},\log(g),A_0)$ at effective temperatures between $3500 \text{ K}<T_{\text{eff}}<8000\text{ K}$, in $10$ K steps, and extinctions between $0<A_0<1.5$, in $0.02$ steps. We analyse the results using $867$, $87$ and $10$ wavelength observations {evenly spaced over the spectra between $200$ nm and $2000$ nm}. We find that the correlation between extinction and effective temperature is largely present across all parameters in this example. {{However, at lower extinctions the impact of the degeneracy on the extinction parameter standard deviation is more pronounced; because, for low extinctions, the extinction curve has a small effect on the long wavelength observations. In Figure \ref{fig:different_ext}, we show the parameter standard deviation calculated from the Fisher information analysis for the effective temperature and the extinction. In particular, we see that the ratio $A_0/\sigma(A_0)$ is lower for low-extinction sources, due to the degeneracy having a more pronounced impact on the size of the extinction parameter standard deviation.}} 

{We perform the same calculation for sources with $A_0<0.2$ mag with $8$ wavelength observations a) ranging across the optical wavelength range, b) ranging across the full spectrum and c) ranging across the infrared wavelength range to inspect the information each portion of the spectra provides information. We display the results in Figure \ref{fig:wl_obs}, where we see the optical observations are the most useful in constraining extinction for small variations at low extinction. At high extinctions, the introduction of longer wavelength observations disentangles the degeneracy between the two parameters, which is a well-known effect and has been noted in \citep{2011MNRAS.411..435B}. The introduction of the near-to-far IR wavelength observations will put a constraint on the large values of extinction that provide a good fit to the data. However, for small values of extinction, the longer wavelength flux observations do not have the same effect in {distinguishing} the difference between extinction and effective temperature. If we know that extinction is low, observations in the optical are far more valuable in constraining the extinction and effective temperature.}} We display synthetic spectra to show how a smaller value of extinction exhibits a larger degeneracy to the effective temperature in Figure \ref{fig:a0_01} than when the extinction is higher (Figure \ref{fig:a0_1}).}


\begin{figure*}
    \centering
    \includegraphics[width=\textwidth]{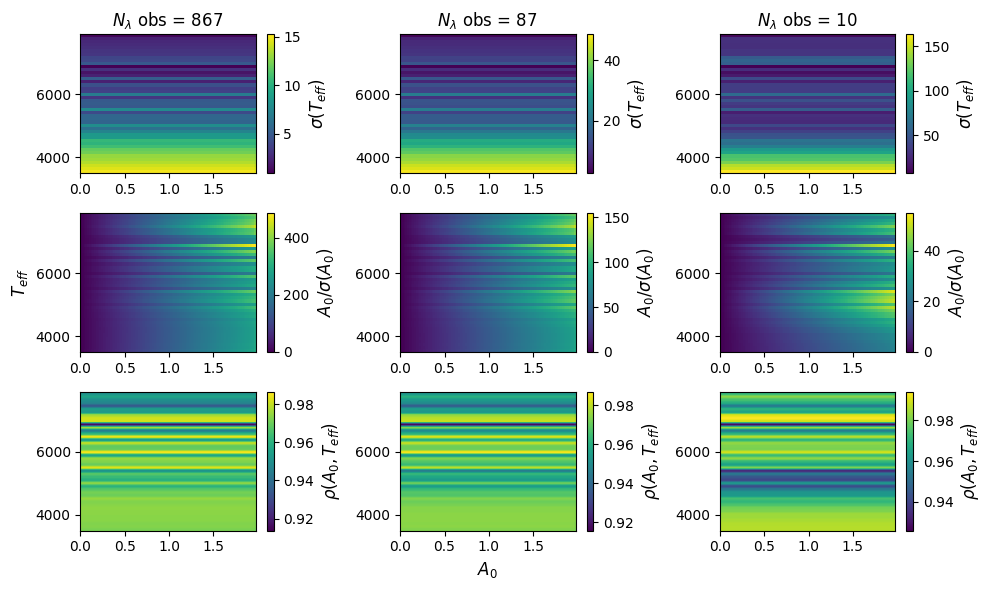}
    \caption{{Precision and degeneracy using a Fisher information analysis of fitting for the effective temperature (K) and extinction (mag). $N_\lambda$ obs indicates the number of evenly spaced wavelength observations one observes from a spectrum. The top row indicates that the standard deviation of an effective temperature maximum likelihood estimate decreases with the number of wavelength observations, but is larger for lower effective temperatures. The middle row shows that the fractional error on the extinction estimate is higher in low-extinction regimes. Moreover, the strong degeneracy between the two parameters is present everywhere (bottom row).}}
    \label{fig:different_ext}
\end{figure*}

\begin{figure}
    \centering
    \includegraphics[width=0.5\textwidth]{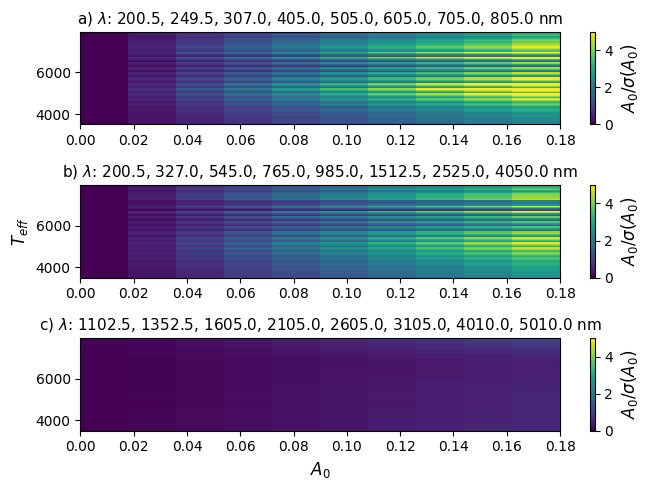}
    \caption{{The ratio $A_0/\sigma(A_0)$ for the maximum likelihood estimate calculated through a Fisher information analysis. We present the ratio as a function of $A_0$ (mag) along the x-axis for $0<A_0<0.2$ mag. The $\lambda$ values in the title of the subplots indicate the wavelength (nm) that observations were taken. We see that in the low-extinction regime, the optical observations provide the most information about $A_0$ for small variation.}}
    \label{fig:wl_obs}
\end{figure}

\begin{figure}
    \centering
    \includegraphics[width=0.5\textwidth]{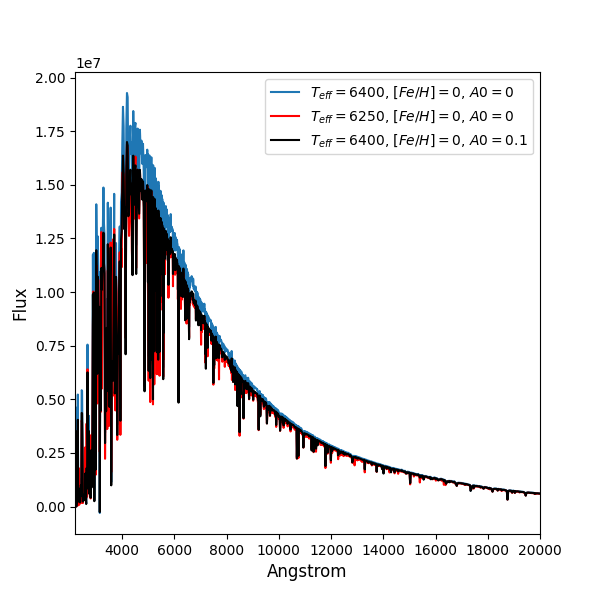}
    \caption{Three solar metallicity SEDs of varying effective temperature and extinction. We see that for a small amount of extinction $A_0=0.1$ mag, we can find a similar SED corresponding to a zero extinction star with a different effective temperature. In particular, when we integrate over passbands this becomes increasingly difficult.}
    \label{fig:a0_01}
\end{figure}
\begin{figure}
    \centering
    \includegraphics[width=0.5\textwidth]{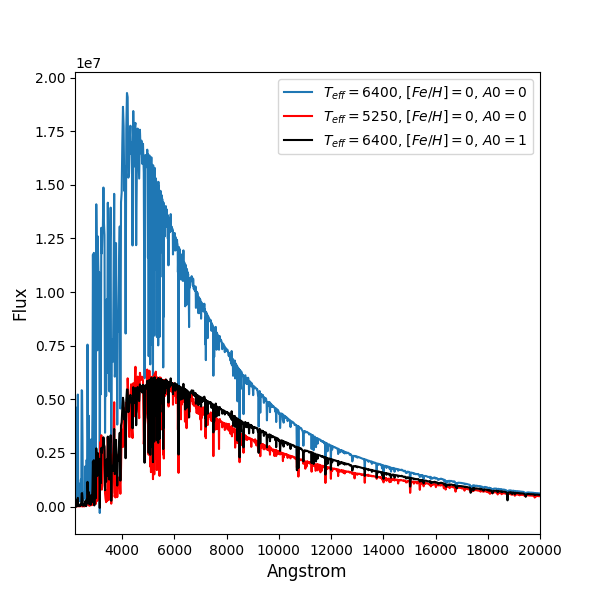}
    \caption{Three solar metallicity SEDs of varying effective temperature and extinction. We see that for a relatively large amount of extinction $A_0=1$ mag, the degeneracy between extinction and effective temperature is not as pronounced. It is easier to detect the difference between the effective temperature and extinction for high extinction regions when one has observations of the flux at longer wavelengths.}
    \label{fig:a0_1}
\end{figure}

\subsection{Uniform Synthetic Sample Generation}\label{uniform}
\par
To validate the region in which we can {precisely} investigate extinction, we generate stars from a large range of stellar parameters {with fixed $A_0=0.15 \, \rm mag$} and see how the extinction posterior distribution depends on the underlying stellar parameters that generated the star. We begin by outlining the procedure to generate the uniform sample of stars and in further sections discuss the results of running our code on the full parameter range.

We generate a sample of low-extinction stars with parameters sampled uniformly over a large range. To begin, we model the positional data and its measurement error as a function of Gaia $G$ magnitude and distance on a high Galactic latitude region by querying the Gaia database. To generate a single star in the synthetic sample we sample a right-ascension and declination value uniformly from an arbitrary four-degree circle on the high-latitude sky and then sample a metallicity, mass, and $\log(\text{age})$ value uniformly over a $[-2,0.5] \, \rm{dex}\times [0,3] \, M_\odot\times [8.9,10.3]$ $\log(\rm years)$ grid. With a value for the metallicity, mass, and $\log(\text{age})$, we can use our forward model to generate consistent, derived stellar parameters and synthetic photometry $\boldsymbol{\hat{x}}$ in the photometric passbands of interest. Not all initial values of metallicity, mass, and $\log(\text{age})$ generate a solution of the stellar models and we reject all sources with invalid solutions. 

Simultaneously, we {draw a distance} for the region of interest using the prior in Equation \ref{bjp} and convert it to a parallax value to be included as data. Moreover, we {construct} a realistic parallax measurement error from the distribution of the region as a function of distance. Once we have a value for the metallicity, mass, $\log(\text{age})$, and position we can generate realistic synthetic measurement error values for estimating each of the stellar parameters {by using our realistic modelled distribution of photometric, spectroscopic and astrometric errors. }

For every data point, we simulate the photometry {and add realistic measurement errors}. We implement this through {setting $\sigma_{\text{floor}}=0.05$ mag and, therefore, ${\Sigma}_{jj}=\sigma_{\text{floor}}^2+\sigma^2_j$ so that each observed passband magnitude} ${\hat{x}}_j$ is sampled from $N(x_j,{{\Sigma}^2_{jj}})$, where $\sigma_j$ is the synthetic measurement error value from the function modelled on the high latitude region defined above. This is to account for any errors that may arise from the definition of the passbands or incorrect modelling from the isochrones. We chose the value of $0.05$ to account for the maximum difference in the photometry provided by the different sets of isochrones and the difference in the extinction provided by using different extinction laws. Choosing a larger value has the effect of making the posterior distribution wider and allowing for a greater range of extinctions in the posterior samples. If we do not allow for this {model error}, the high accuracy of the Gaia photometry dominates the likelihood function and forces the model into a solution which is false due to the systematic error of the theoretical models being greater than the measurement error of the Gaia photometry. In Section \ref{floor_error}, we inspect the results of assuming the models match observations perfectly within the measurement error.

{To simulate the effect of uncertainties on spectroscopic estimates of the effective temperature and the metallicity we perturb the values }supplied by the theoretical isochrones by {drawing them} from normal distributions $N(T_{\text{eff}},(110 \, \rm{K})^2)$, and  $N([Fe/H],(0.1\,\rm{dex})^2)$, respectively. Finally, we use the \cite{f99} extinction law with $R_V=3.1$ and $A_0=0.15 \,\rm{mag}$ to add extinction to the magnitudes in each passband. {Our model incorporates a more flexible prior on the $R_V$ parameter to allow for values normally distributed around 3.1 but we note that this has a small effect on the final posterior and we generate samples using a fixed value of $R_V$. }

We continually sample until we have $10000$ stars sampled uniformly across the metallicity, mass and $\log(\text{age})$ parameter space. We recall that the widths of the spectroscopic {constraint on the parameter} are in line with \cite{Andrae_2023}, except the effective temperature which has been significantly increased to account for the large degeneracy between effective temperature and extinction.

\begin{table*}
  \centering
  \begin{tabular}{||c c c c||}
    \hline
    \textbf{Data Set} & \textbf{Median $\Delta A_0$}& \textbf{Median $\Delta A_0$ ($[Fe/H]>-0.3$)} & \textbf{Median Std$(A_0)$} \\
    \hline
    \hline
     Uniform prior, $\sigma_{\text{floor}}=0$ mag& $0.075$ mag& $0.168$ mag& $0.113$ mag\\
     \hline
    Uniform prior, $\sigma_{\text{floor}}=0.05$ mag & $0.206$ mag& $0.461$ mag&$0.209$ mag\\
    \hline
    $[Fe/H]$ prior, $\sigma_{\text{floor}}=0.05$ mag& $0.058$ mag & $0.291$ mag& $0.159$ mag\\
    \hline
    $T_{\text{eff}}$ constraint, $\sigma_{\text{floor}}=0.05$ mag & $0.022$ mag & $0.076$ mag& $0.075$ mag\\

    \hline
    \hline
  \end{tabular}
  \captionsetup{justification=centering} 
  \caption{Median difference between the posterior mean extinction and the true extinction value of $A_0=0.15$ mag, with the median taken over the full synthetic sample. We inspect the accuracy of recovering the extinction {for zero and nonzero values of  $\sigma_{\text{floor}}$}. Moreover, see that different spectroscopic constraints on the parameters will tighten the extinction posterior around the true value to a differing extent.}
  \label{tab:widths}
\end{table*}

\subsection{Parameter Sensitivity and Degeneracy: Uniform Priors on $[Fe/H]$ and $A_0$}\label{uniform_priors}

Using the generated synthetic samples, we assess the model's ability to consistently recover input parameters over a uniform grid. In this section, {we alter the $[Fe/H]$ and $A_0$ priors of the model} originally defined in Section \ref{methods}. We run the nested sampling algorithm to generate samples of the posterior over the entire parameter space and investigate the sensitivity of the likelihood to each parameter. We test the case where the model exhibits {an internal error} in the photometry of  $\sigma_{\text{floor}}=0.05$ mag in this section, and the case when the model is perfect ($\sigma_{\text{floor}}=0.0$ mag ) in Section \ref{floor_error}. We still use the IMF as an initial mass prior and use the distance prior defined in Section \ref{methods}.

{For a single star, we define the posterior standard deviation for a parameter $\theta$ as $\sigma(\theta)=\sqrt{\text{Var}(\theta)}$, where the variance is calculated from the posterior samples. Moreover, we define $\Delta(\theta)$ to be the difference between the posterior mean value of $\theta$ and the true value. Assume we have a set of $n$ stars each with posterior samples from their individual posteriors; we define the median posterior standard deviation for a parameter $\theta$ across the sample as the median value of $\sigma(\theta)^{(i)}$, the individual posterior standard deviation for each star indexed by $(i)$, taken over all sample stars $i=1,...,n$. We define the median posterior mean of $\theta$ across the sample similarly.}

We recall that each star in the synthetic sample has a true extinction of $A_0=0.15$. Under uniform prior assumptions on  $[Fe/H]$ and $A_0$, the model performs poorly in retrieving an accurate point estimate of the model parameters. {In Figure \ref{fig:corner_uniform} we display a corner plot of the posterior distribution for a single star with  $[Fe/H]=-0.125$ dex, $T_{\text{eff}}=5916$ K, age=$10^9$ yrs, $M_G=4.7$ mag, $d=800$ pc, $A_0=0.15 \,\rm{mag}$ and $m=M_\odot$.} We calculate the posterior distribution for each star in our sample. Through comparing the posterior samples with the true synthetic value, we find {that} the median difference between the posterior mean extinction and the true extinction is $0.21 \,\rm{mag}$ and the median posterior extinction standard deviation across the sample is $0.209 \,\rm{mag}$. The median difference between the posterior mean effective temperature and the true effective temperature is $428.65$ K and the median effective temperature standard deviation across the sample is $346.73$ K. Finally, the median difference between the posterior mean metallicity and the true metallicity is $-0.30$ dex with a standard deviation of $0.52$ dex. In Table \ref{tab:widths}, we illustrate the extinction posterior median and standard deviations for different cases. 

We display the standard deviation, derived from the posterior samples, of the effective temperature, log surface gravity and each of the model parameters in Figure \ref{fig:sensitivity_wfloor}. From the posterior samples, we derive the correlation between extinction and the effective temperature, log surface gravity and other model parameters, respectively, which we display in Figure \ref{fig:corr_wfloor}. We see that the model {cannot constrain strongly} each of the parameters under the assumed errors. In particular, the effective temperature is increasingly hard to constrain at brighter absolute Gaia $G$ magnitudes, $M_G$. Moreover, the posterior standard deviation on the extinction parameter is too high to use a point estimate value to constrain the true extinction. This arises because of the significant degeneracies {among} extinction, the effective temperature, and metallicity. 

\begin{figure*}
    \centering
    \includegraphics[width=\textwidth]{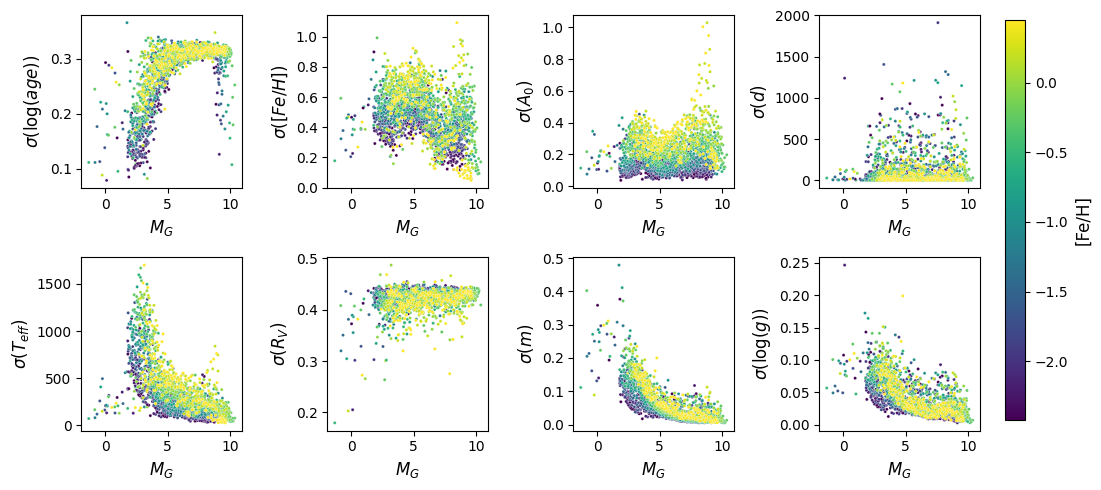}
        \caption{Posterior standard deviation, assuming uniform priors on  $[Fe/H]$ and $A_0$, of the effective temperature (K), log surface gravity (dex) and of each of the model parameters for the uniform sample of stars with $\sigma_{\text{floor}}=0.05$ mag. $M_G$ refers to Gaia absolute $G$ magnitudes. We see that when using uniform priors and assuming a nonzero value of $\sigma_{\text{floor}}$, the model is not sensitive to extinction. The true value of extinction in this sample is $A_0=0.15\,\rm{mag}$, which is of the same order as the standard deviation of extinction. $d$ is in parsecs, $m$ is a multiple of $M_\odot$ and $\log(\rm{age})$ is in $\log(\rm{yrs}).$}
    \label{fig:sensitivity_wfloor}
\end{figure*}

\begin{figure*}
    \centering
    \includegraphics[width=\textwidth]{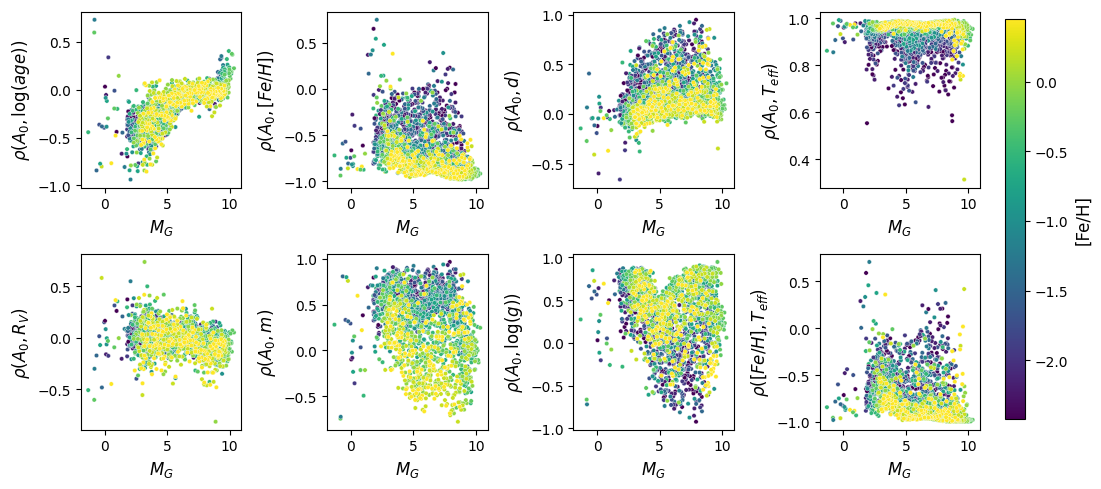}
        \caption{Parameter correlations from posterior samples of each star, assuming uniform priors on $[Fe/H]$ and $A_0$, for the uniform sample of stars with $\sigma_{\text{floor}}=0.05$ mag. We display the correlation between extinction (mag) and the effective temperature (K), log surface gravity (dex) and other model parameters, respectively. $M_G$ refers to Gaia absolute $G$ magnitudes. Moreover, in the bottom-right plot, we display the correlation between metallicity and effective temperature. We see that when using uniform priors and assuming a nonzero value of $\sigma_{\text{floor}}$ there are significant degeneracies between the extinction, metallicity (dex) and the effective temperature. These degeneracies are inherent to the likelihood and some constraint on one of these parameters is necessary to reduce the width of the extinction posterior. $d$ is in parsecs, $m$ is a multiple of $M_\odot$ and $\log(\rm{age})$ is in $\log(\rm{yrs}).$}
    \label{fig:corr_wfloor}
\end{figure*}

\subsection{Stellar Model Inaccuracy Analysis}\label{floor_error}
There are systematic effects which cause stellar models to mismatch observation for a given input set of stellar parameters. We have identified a region along the main sequence of the HR diagram where stellar models are in the best agreement. In this section, we wish to show how the assumed $\sigma_{\text{floor}}$ defined in our model propagates to the final extinction posterior. {We recall  $\sigma_{\text{floor}}$ is an error term included in quadrature to the measurement error to account for possible systematics in the output photometry and it defaults to $0.05$ mag.} {We set $\sigma_{\text{floor}}=0.0$ mag so that the only noise added to the photometric output
of the stellar models is the modelled error in each passband and repeat the experiment on the uniform sample of stars.}

When comparing point estimates from the derived posterior with the true synthetic values, we find a median difference between the posterior mean extinction and the true extinction is $0.075 \,\rm{mag}$ and the median posterior standard deviation for the extinction parameter across the sample is $0.113 \,\rm{mag}$. The median difference between the posterior mean effective temperature and the true effective temperature is $166$ K and the median effective temperature standard deviation across the sample is $284$ K. Finally, the sample median difference between the posterior mean metallicity and the true metallicity is $-0.17$ dex with a standard deviation of $0.43$ dex. In Table \ref{tab:widths}, we illustrate the extinction posterior median and standard deviations for different cases.

In Figure \ref{fig:comparison_floor_std}, we show the difference in the posterior standard deviation for each star when the data was generated using $\sigma_{\text{floor}}=0.05$ mag (vertical axes) and when we set  $\sigma_{\text{floor}}=0.0$ mag in the synthetic data (horizontal axes). The contours indicate the relative density of points in the diagram. We see that  $\sigma_{\text{floor}}$  has a significant effect on the posterior standard deviation of each parameter. In particular, including a nonzero  $\sigma_{\text{floor}}$  in the analysis has widened the extinction posterior by a significant amount. We do not display the correlations of each parameter with extinction as they are unchanged, except for the extinction-effective temperature degeneracy. {Increasing $\sigma_{\rm floor}$ increases the error on estimating each of the extinction, the effective temperature and the metallicity parameters due to the degeneracy among them}.

\begin{figure*}
    \centering
    \includegraphics[width=\textwidth]{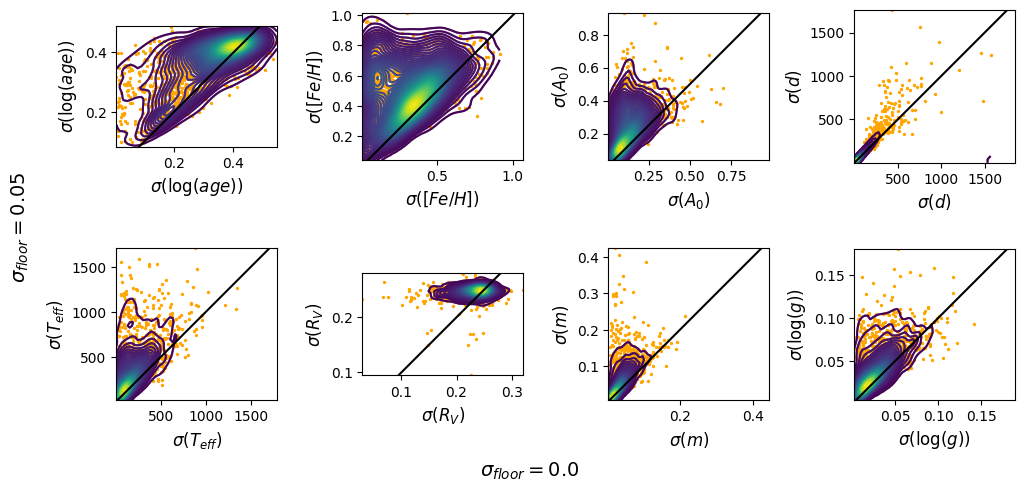}
        \caption{Standard deviation of each parameter derived from the posterior samples when we set $\sigma_{\text{floor}}=0.05$ mag in the synthetic data (vertical axes) and when set  $\sigma_{\text{floor}}=0.0$ mag in the synthetic data (horizontal axes). We overlay the density information on the scatter plot in each case and include the relation $y=x$ as a black line. We see that a nonzero  $\sigma_{\text{floor}}$ will significantly affect the posterior width of the extinction. $A_0$ is in mag, $d$ in parsecs, $m$ is a multiple of $M_\odot$, $[Fe/H]$ in dex, $T_{\rm eff}$ in K, and $\log(\rm{age})$ is in $\log(\rm{yrs}).$}
    \label{fig:comparison_floor_std}
\end{figure*}

\subsection{Parameter Sensitivity and Degeneracy: Spectroscopic Effective Temperature Constraint}\label{teff_constraint}
We show how incorporating a spectroscopic constraint on the effective temperature greatly reduces the width of the extinction posterior. Moreover, we show the extent of the reduction depends on what part of the HR diagram we inspect. We stress that due to the strong degeneracy between extinction and the effective temperature, {a point estimate of one of these parameters within a given error without reference to the other parameter is misleading and should be avoided.} Throughout we assume $\sigma_{\text{floor}} =0.05$ mag.

We run our model on the synthetic uniform sample of stars, assuming that we have a Gaussian constraint on the effective temperature with standard deviation $\sigma_T=110$ K. We choose this value as many current stellar parameter studies which derive effective temperatures from the Gaia XP spectra quote mean absolute differences less than this value. Moreover, when we cross-match and compare the \cite{Andrae_2023} effective temperatures to the highly accurate Gaia ESO \citep{2022ges} effective temperatures for a region at high Galactic latitudes we find a mean difference of $110$ K. {In Figure \ref{fig:corner_uniform_except_metal_teff_cont} we display a corner plot of the posterior distribution for a single star with  $[Fe/H]=-0.125$ dex, $T_{\text{eff}}=5916$ K, age=$10^9$ yrs, $M_G=4.7$ mag, $d=800$ pc, $A_0=0.15\,\rm{mag}$ and $m=M_\odot$.} We generate a posterior distribution for all stars in our sample. Through comparing the posterior samples with the true synthetic value, we find} a median difference between the posterior mean extinction and the true extinction of $0.022 \,\rm{mag}$ and the median extinction posterior standard deviation across the sample is $0.075\,\rm{mag}$. The median difference between the posterior mean effective temperature and the true effective temperature is $27.35$ K and the median effective temperature standard deviation across the sample is $87.00$ K. Finally, the sample median difference between the posterior mean metallicity and the true metallicity is $-0.003$ dex with a standard deviation of $0.368$ dex. In Table \ref{tab:widths}, we illustrate the extinction posterior median and standard deviations for different cases.

\begin{figure*}
    \centering
    \includegraphics[width=\textwidth]{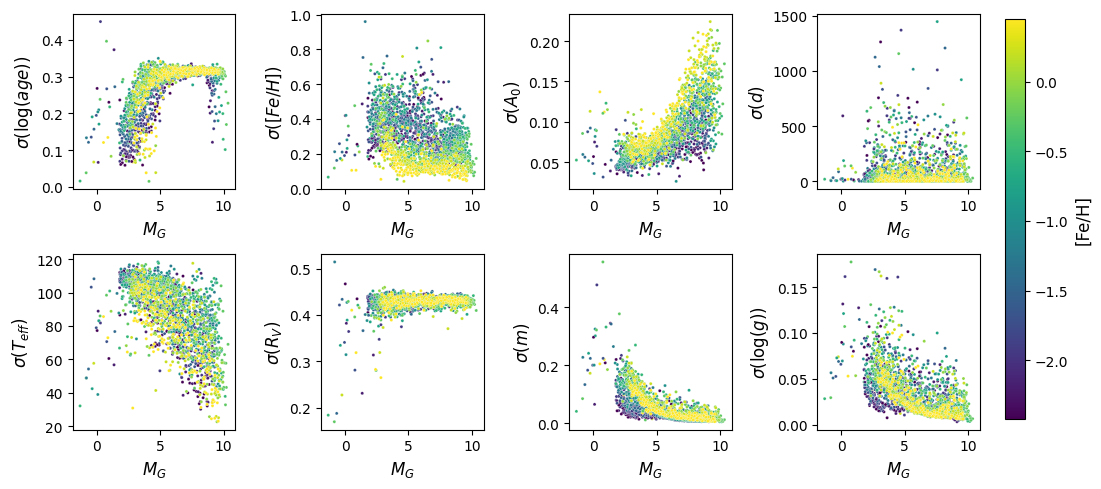}
        \caption{Posterior standard deviations assuming uniform priors, $\sigma_{\text{floor}} =0.05$ mag, and an effective temperature Gaussian constraint with standard deviation $\sigma_T=110$ K. We display the standard deviation of the effective temperature (K), log surface gravity (dex) and of each of the model parameters for the uniform sample of stars. $M_G$ refers to Gaia absolute $G$ magnitudes. We see that when using uniform priors and a spectroscopic constraint on the effective temperature, the model is far more sensitive to extinction. The true value of extinction in this sample is $A_0=0.15 \,\rm{mag}$. We see the posterior extinction standard deviation (mag) is reduced for brighter absolute Gaia magnitudes along the main sequence. $d$ is in parsecs, $m$ is a multiple of $M_\odot$ and $\log(\rm{age})$ is in $\log(\rm{yrs}).$}
    \label{fig:constraint_teff}
\end{figure*}

\begin{figure*}
    \centering
    \includegraphics[width=\textwidth]{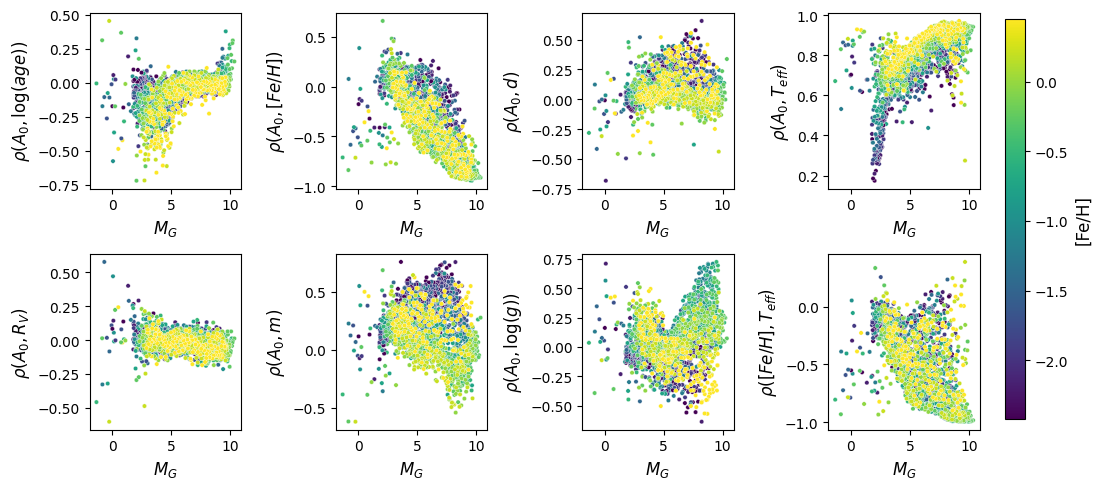}
        \caption{Parameter correlations from posterior samples of each star, assuming uniform priors and an effective temperature Gaussian constraint with standard deviation $\sigma_T=110$ K. We display the correlation between extinction (mag) and the effective temperature (K), log surface gravity (dex) and other model parameters, respectively. $M_G$ refers to Gaia absolute $G$ magnitudes. Moreover, in the bottom-right plot, we display the correlation between metallicity (dex) and effective temperature. We see that when using uniform priors and having a spectroscopic constraint on the effective temperature significantly affects the degeneracy between model parameters. This will directly reduce the width of the posterior extinction distribution. $d$ is in parsecs, $m$ is a multiple of $M_\odot$ and $\log(\rm{age})$ is in $\log(\rm{yrs}).$}
    \label{fig:corr_teff}
\end{figure*}

We display the standard deviation, derived from the posterior samples, of the effective temperature, log surface gravity and each of the model parameters in Figure \ref{fig:constraint_teff}. From the posterior samples, we derive the correlation between extinction and the effective temperature, log surface gravity and other model parameters, respectively, which we display in Figure \ref{fig:corr_teff}. We see that the model is far more sensitive to each of the parameters under the assumed errors when we have a constraint on the effective temperature. We noted before that the effective temperatures were increasingly hard to constrain at brighter absolute Gaia $G$ magnitudes {in the absence of spectroscopy}. Thus, constraining this value will greatly constrain the extinction in the same region of the HR diagram, due to the strong degeneracy between the parameters. We also see that at brighter absolute magnitudes the constraint on the effective temperature will reduce the degeneracy between extinction and the other parameters, in particular, the metallicity. However, the degeneracy at fainter absolute magnitudes is still strong between metallicity and extinction.

\begin{figure*}
    \centering
    \includegraphics[width=\textwidth]{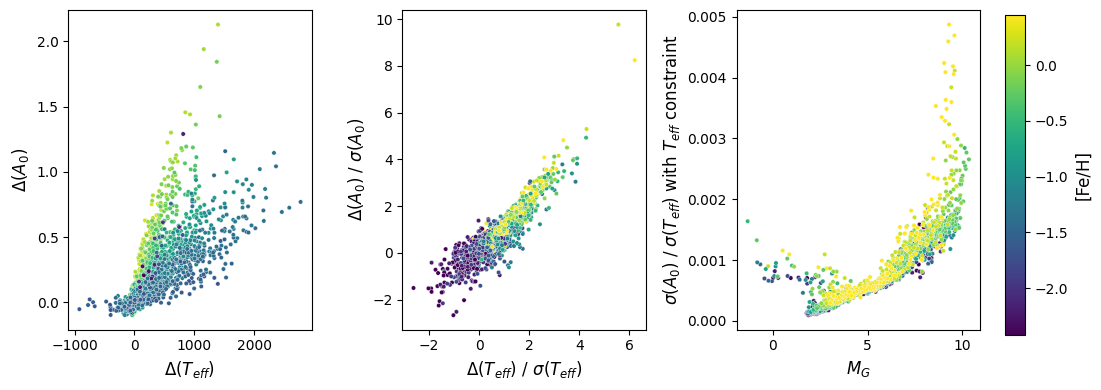}
        \caption{The sensitivity between constraining the extinction parameter (mag) and the effective temperature (K) using uniform priors. $\Delta(A_0)$ (mag) and $\Delta(T{\text{eff}})$ (K) represent the difference between the mean extinction/effective temperature and the true extinction/effective temperature, respectively. $\sigma(A_0)$ and $\sigma(T_{\text{eff}})$ are the posterior standard deviation for the extinction and effective temperature parameter respectively. We find that introducing a spectroscopic constraint on the effective temperature will leave the centre plot invariant. Thus, we can inspect the sensitivity of the effective temperature constraint needed to give a respective constraint on extinction in the right-most plot.}
    \label{fig:constraint_orig}
\end{figure*}

We find that the ratio between $\Delta(A_0)/\sigma(A_0)$ and $\Delta(T_{\text{eff}})/\sigma(T_{\text{eff}})$ is constant for all realistic spectroscopic constraints on the effective temperature ($\sigma_T>5 K)$. This allows us to demonstrate a rough (but good) approximation of propagating the spectroscopic constraint of the effective temperature to the posterior width of the extinction parameter. We display this in the right plot of Figure \ref{fig:constraint_orig}.

\subsection{Parameter Sensitivity and Degeneracy: Astrophysical Metallicity Prior}\label{metallicity_prior}

We now illustrate the results of fitting the model to the data when we use the prior defined in Section \ref{a_informed_p}, which modelled the LAMOST metallicities as a function of distance for stars at high Galactic latitudes. {We consider all stars within 2000pc. {We display the metallicity prior in the top four images in Figure \ref{fig:feh_prior_data}.}} Our previous analysis shows that constraints on either the effective temperature or the metallicity are necessary to constrain the extinction without adding any prior on the extinction parameter of our model, particularly when $\sigma_{\text{floor}}>0.0$. We fit our model to the uniform synthetic sample with  $\sigma_{\text{floor}}= 0.05$ mag.

We find that the prior on the metallicity significantly improves our ability to constrain the extinction parameter across the whole sample, particularly for stars which have metallicities that are highly likely under our prior assumption. {In Figure \ref{fig:corner_uniform_except_metal} we display a corner plot of the posterior distribution for a single star with  $[Fe/H]=-0.125$ dex, $T_{\text{eff}}=5916$ K, age=$10^9$ yrs, $M_G=4.7$ mag, $d=800$ pc, $A_0=0.15 \,\rm{mag}$ and $m=M_\odot$.} Through generating point estimates for the parameters of each star using their respective posteriors, we find {that} the median difference between the posterior mean extinction and the true extinction is $0.058 \,\rm{mag}$ and the median posterior standard deviation is $0.159$. We do not include a tight spectroscopic constraint on the metallicity as a Gaussian term in the likelihood function using a value from spectroscopic surveys due to it being difficult to calibrate spectroscopic metallicities to those from stellar evolution code. We do not wish to introduce any unknown systematic error when training to constrain values of extinction, thus, we incorporate our spectroscopic knowledge of the metallicity via an astrophysical prior. 

\begin{figure*}
    \centering
    \includegraphics[width=\textwidth]{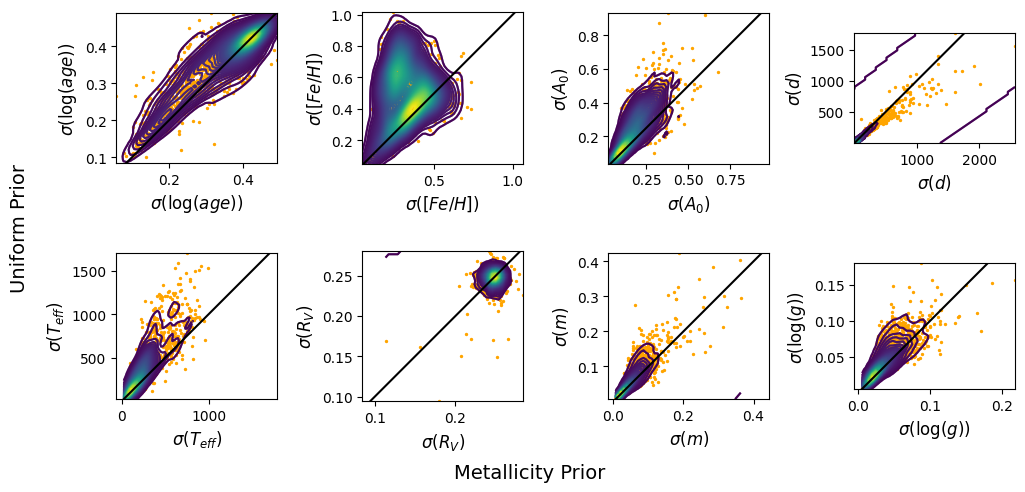}
        \caption{Parameter standard deviations from posterior samples of each star with $\sigma_{\text{floor}}=0.05$ mag on the photometry when (vertical axes) assuming a uniform prior on the metallicity versus (horizontal axes) using a metallicity prior derived from LAMOST metallicities at high Galactic latitudes. We see that the metallicity prior is useful in constraining the extinction, metallicity and effective temperature.}
    \label{fig:feh_std}
\end{figure*}

We display the posterior standard deviation of the effective temperature, log surface gravity and each of the model parameters in Figure \ref{fig:feh_std}. We see that the model is far more sensitive to each of the parameters under the assumed errors when we have {an astrophysical metallicity prior}. In particular, the metallicity, extinction, and effective temperature posterior standard deviations are reduced over the sample. We omit an illustration of the correlations due to them being {similar to} Figure \ref{fig:corr_wfloor}.

\begin{figure}
    \centering
    \includegraphics[width=0.5\textwidth]{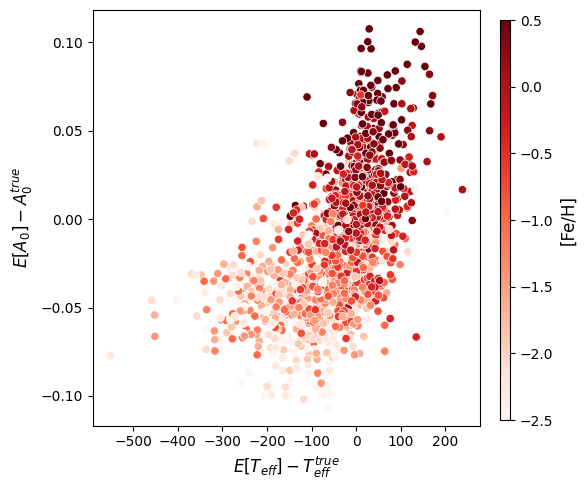}
        \caption{Comparing the difference between our posterior mean extinction and the true extinction  (mag) for each star against the difference between the posterior mean effective temperature and the true effective temperature (K), for the uniform sample using a metallicity prior. The relative occurrence of points in this diagram is not relevant as the samples have been generated to be uniform over the metallicity prior bounds. An informed prior on the metallicity as a function of distance is essential to mitigate against this bias.}
    \label{fig:prior_feh}
\end{figure}

Due to the degeneracies between metallicity and other parameters, our model is not particularly sensitive to metallicity using just the Gaia, 2MASS and ALLWISE photometric passbands. Sources with a true, low prior probability metallicity will return an inaccurate metallicity point estimate and bias the effective temperature and extinction point estimates to an inaccurate value. We can see this effect by comparing the difference between our posterior mean extinction and the true extinction for each star against the difference between the posterior mean effective temperature and the true effective temperature. We display a plot of this comparison in Figure \ref{fig:prior_feh}, where outside the metallicity prior we have increasingly bad point estimates for extinction.

More flux observations in the optical wavelength range allow one to constrain the metallicity better for such stars. Constraining the metallicity reliably can be contaminated by systematics from the stellar models. In future work, we will look at accurately calibrating the Gaia BP/RP spectra to be used as spectral observations in our model. This will allow us to further constrain the metallicity and provide more robust extinction estimates, particularly for outlier sources which lie outside the bulk of the metallicity prior probability.

\subsection{A Family of Equally Likely Extinction-Effective Temperature Pairs}\label{transform}

\begin{figure*}
    \centering
    \includegraphics[width=\textwidth]{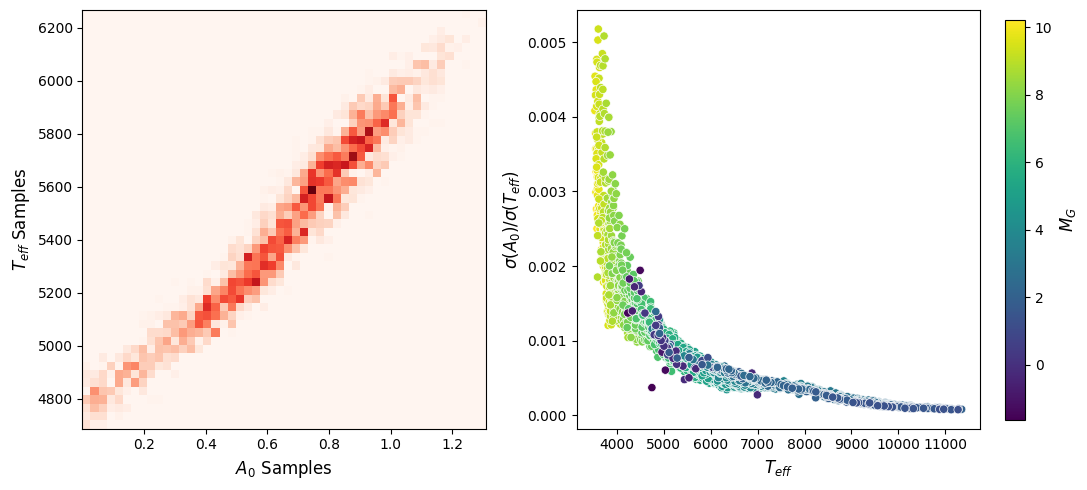}
        \caption{{Left plot: Posterior samples of the effective temperature (K) and extinction parameter (mag) for {all of the stars from the uniform sample as described in Section \ref{uniform} displayed in a single plot together to highlight the degeneracy}. {The synthetic stars used to generate this plot have an extinction value of $A_0=0.15 \,\rm{mag}$.} This illustrates the degeneracy over the entire data set and allows us to visualise the strong correlation as displayed in the top-right plot of Figure \ref{fig:corr_wfloor}.} Right plot: the ratio, $\sigma(A_0)/\sigma(T_{\text{eff}})$, for different values of $T_{\text{eff}}$. The hue indicates the absolute Gaia $G$ magnitude of the point. We note that this ratio is {independent} of prior assumptions on the metallicity, effective temperature and extinction unless they are particularly narrow. In particular, constraining the metallicity will reduce the {scatter} of the curve.}
    \label{fig:constraint}
\end{figure*}

The strong degeneracy between extinction and effective temperature allows for a large family of point estimates which give a similar posterior probability under the assumed model errors. Introducing astrophysical priors and spectroscopic constraints only restricts our posterior to a subset of this family. {We display the posterior samples of the effective temperature and extinction for {all of the stars from the experiment in Section \ref{uniform} {in the left image of Figure \ref{fig:constraint} to highlight the degeneracy. The synthetic stars used to generate this plot have an extinction value of $A_0=0.15 \,\rm{mag}$.} We assume a uniform metallicity and extinction prior as per Section \ref{uniform_p_sec}, and  $\sigma_{\text{floor}}=0.05$ mag.}}

We find that the ratio $\sigma(A_0)/\sigma(T_{\text{eff}})$ can be described as a function of the effective temperature and it is {independent} of the prior distributions on the stellar parameters. The introduction of priors (provided they are not extremely narrow) does not remove the degeneracy but restricts the degeneracy to a subset of the parameters. We display this relation in the right plot of Figure \ref{fig:constraint}, and note the ranges of effective temperatures in both images. Unconstrained by priors, the range of effective temperatures and extinctions is approximately given by the length of the intersection between the line passing through the parallax-corrected photometric point in the direction of the extinction vector, and isochrone manifold in photometric space. This intersection depends on the model's errors; higher errors mean more points on the manifold will appear to intercept the line. 

We can use this relation to 'transform' extinction into effective temperature and vice versa. If we have an effective temperature and extinction pair, $(T_{\text{eff}},A_0)$, we can convert a small amount of extinction, $\delta A_0$, into a small change of effective temperature via the relation $\delta T_{\text{eff}}=-\delta A_0 (\sigma(T_{\text{eff}})/\sigma(A_0))$, by using $\sigma(A_0)/\sigma(T_{\text{eff}})$ to move along the curve in the right plot of Figure \ref{fig:constraint}. Moreover, we can perform this operation repeatedly to move along the degeneracy and generate a family of pairs with similar posterior probability, provided we do not move suitably far away from the posterior mode. This is useful when comparing results against spectroscopic surveys.

\section{Results and Model Validation With Real Data}\label{validation_real}
We {display results and} further validate aspects of our model using real photometric, parallax and spectroscopic data. We illustrate how sensitive the extinction posterior is to assuming knowledge of the extinction parameter or a spectroscopic {constraint} of the effective temperature. 

\subsection{Comparison with LAMOST Derived Parameters}\label{crossmatch_LA}
In this section, we validate our model against the LAMOST DR8 \citep{LAMOSTDR8} {spectroscopic estimates of stellar parameters} which have been derived using the LASP stellar parameter model \citep{2014IAUS..306..340W}. We note that for high Galactic {latitude} stars, both LASP and the LAMOST Spectrograph Response Curves \citep{LAMOST_response} ignore the effects of interstellar extinction when deriving stellar parameters. We will show that our mean posterior effective temperatures match the LAMOST effective temperatures when we convert our mean extinction into an effective temperature term through the transformation along the extinction-effective temperature degeneracy, introduced in Section \ref{transform}.

We select sources from LAMOST with Galactic latitude $|b|>60^\circ$ and cross-match with Gaia DR3, 2MASS and ALLWISE as outlined in Section \ref{Data}. We remove all sources with LAMOST effective temperature error greater than $50$ K to match the mean error in effective temperature quoted by \cite{Andrae_2023}. For the same reason, we remove all sources which exhibit a metallicity error greater than $0.1$ dex. Finally, we remove all sources with distance from \cite{Bailer_Jones_distance} exceeding $2$ kpc to mitigate against the effects of parallax error. 

We fit our model to the photometry and parallax data assuming a prior extinction distribution {with exponential scale parameter $d_A=0.2$ mag}. {We do not use spectroscopic estimates of stellar parameters in the likelihood function} and we find that there is a systematic offset between our posterior mean effective temperatures and those derived by LAMOST with a median difference in the sample of $223.12$ K. A scatter plot of the LAMOST effective temperatures and the posterior mean effective temperatures for each star in the sample is displayed in the left plot of Figure \ref{fig:compare_lamost}. The median value of the standard deviations of the effective temperature across the sample is $173.90$ K. The median difference between the posterior metallicity and the LAMOST-derived metallicity is $-0.02$ dex, and the median value of the standard deviations of the posterior metallicity across the sample is $0.20$ dex. The median difference between the log surface gravity derived from the posterior and LAMOST's values is $0.05$ dex, and the median posterior standard deviation across the sample is $0.04$ dex. {If we fit our model to the data but assume that the extinction is $A_0=0 \,\rm mag$, then we find {that} the systematic offset between the posterior effective temperatures and the LAMOST derived values is diminished.}

\subsection{Extinction-Corrected Effective Temperatures to Match LAMOST }

The systematic differences between our values and LAMOST's can be misleading due to the strong degeneracy between extinction and the effective temperature of a star. {We find {that} the difference is significantly reduced if we assume $A_0=0 \, \rm mag$ in our model fit. This is not surprising as LAMOST fixes the extinction to zero in their derivation of the effective temperature.}

We recall that the two parameters are intimately related via the curve in the right plot of Figure \ref{fig:constraint}. In our model, we have allowed for stars to have non-zero extinction, this assumption means that the effective temperature has not been fixed by the extinction assumption but is restricted to a region dictated by the prior on $A_0$. Therefore, the point estimate of the effective temperature does not represent the value you would obtain if you assumed zero extinction, but has been restricted to a range dictated by the extinction prior and the data. 

\begin{figure*}
    \centering
    \includegraphics[width=\textwidth]{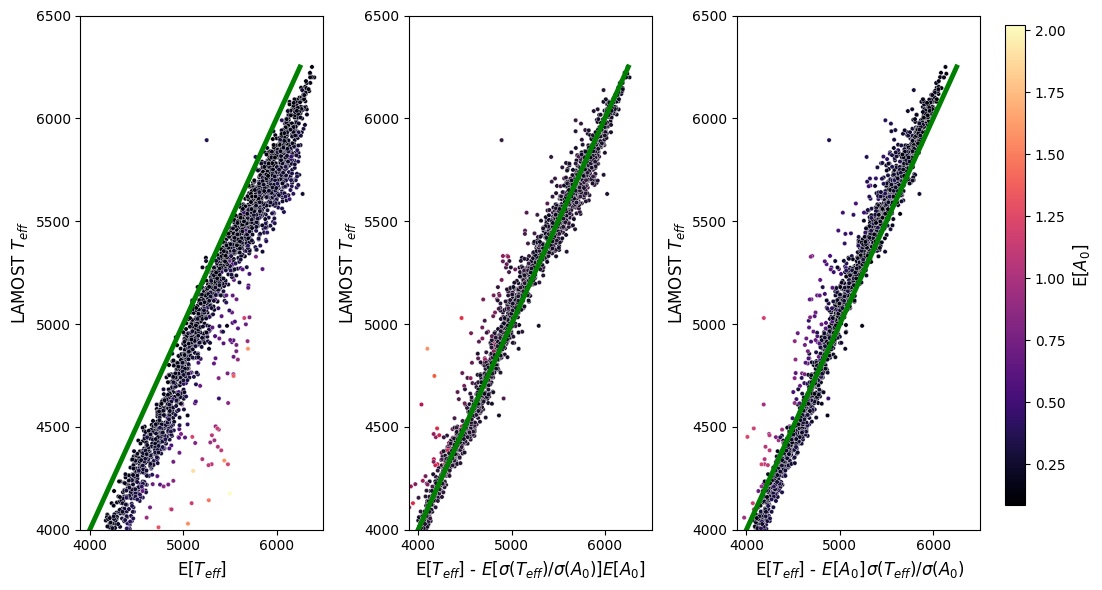}
    \caption{Comparison of posterior mean effective temperatures (K) with LAMOST illustrating that we can reproduce the LAMOST effective temperatures. In the left-most plot, we see the LAMOST effective temperature values against the posterior mean effective temperatures for each star in the sample. By converting the posterior mean extinction for each star into effective temperature we can recover the effective temperatures derived by assuming no extinction. The middle plot shows the correction by applying the uniform transformation and in the right-most plot we make the transformation using the relation from Figure \ref{fig:constraint}. {In the middle plot we calculate ${E}[\sigma(T_{\rm teff})/\sigma(A_0)]$ by taking the average value of $\sigma(T_{\rm teff})/\sigma(A_0)$ from the LAMOST sample. We note that the values of $\sigma(T_{\rm teff})/\sigma(A_0)$ from the LAMOST sample are consistent with the curve in the right plot of Figure \ref{fig:constraint}.}}
    \label{fig:compare_lamost}
\end{figure*}

Therefore, when taking mean values of the posterior we expect our effective temperatures to be systematically different. Through the relation in Figure \ref{fig:constraint}, we can transform extinction values into effective temperature along the vector of degeneracy. Here, we wish to transform all of the extinction for each star into effective temperature. Thus, we transform our posterior mean effective temperature as $E[T_{\text{eff}}]\mapsto E[T_{\text{eff}}]-E[A_{0}]\frac{\sigma(T_{\text{eff}})}{\sigma(A_0)}$, or like $E[T_{\text{eff}}]\mapsto E[T_{\text{eff}}]-E[A_{0}]E[\frac{\sigma(T_{\text{eff}})}{\sigma(A_0)}]$ as an approximation if we wish to use a constant value of $\sigma(T_{\text{eff}})/\sigma(A_0)$ (see Section \ref{transform} for a background to this transformation). Each of the expectations is taken over the posterior for each star. We display a plot of LAMOST's derived effective temperatures together with our derived posterior mean effective temperatures, and the corrected temperatures, respectively, in Figure \ref{fig:compare_lamost}.

In the left-most plot of Figure \ref{fig:compare_lamost} we see the LAMOST effective temperature values against the posterior mean effective temperatures for each star in the sample. There is a clear systematic difference between the two sets of effective temperatures, but we note that it is not uniform across all temperatures. In the middle and right-most plot, we see where this systematic arises from. By converting the posterior mean extinction for each star into effective temperature we can recover the effective temperatures derived by assuming no extinction. The middle plot shows the correction by applying the uniform transformation $E[T_{\text{eff}}]\mapsto E[T_{\text{eff}}]-E[A_{0}]E[\frac{\sigma(T_{\text{eff}})}{\sigma(A_0)}]$ {(where ${E}[\sigma(T_{\rm teff})/\sigma(A_0)]$ is the average of $\sigma(T_{\rm teff})/\sigma(A_0)$ from over the sample)} and we see a strong correlation between the two sets of effective temperatures. In the right-most plot, we transform by assuming the value of $\frac{\sigma(T_{\text{eff}})}{\sigma(A_0)}$ from Figure \ref{fig:constraint}, where we also see a strong correlation.

The correct transformation to make is the one which accounts for what part of the HR diagram one is probing (right plot). We can see that this transformation will accurately reconstruct the LAMOST effective temperatures. However, it introduces a slight curve around the line $y=x$. We stress that by introducing an extinction prior, one expects these systematic differences in posterior mean effective temperatures as the width of the posterior depends on which part of the HR diagram one probes. The effective temperature and the extinction parameter are intimately related and one can only compare effective temperatures reliably as point estimates when a good understanding of the extinction prior is known and accounted for. This systematic bending effect, while slight here, has been seen before (see \cite{2023A&A...670A.107H}) in a more pronounced fashion, particularly when comparing the effective temperatures derived from the APOGEE survey with others. We will illustrate in Appendix \ref{apogee_appendix} that one can derive higher correlations between LAMOST and APOGEE using this correction. However, it is important to note that these corrections are mathematical and all we do is choose a set of parameters which achieve similar posterior probability.

 \subsection{Model Validation Using Gaia ESO}

In the previous section, we showed that we could reliably reproduce the LAMOST effective temperature values if we assumed that there was no extinction in the sample. This was because the LAMOST effective temperatures were derived by assuming there is no extinction for high Galactic latitude stars when deriving stellar parameter estimates.

In this section, we wish to use stellar parameters which have been derived without assuming zero extinction and using methods independent of ours. We use the highly accurate spectroscopic features from Gaia ESO (GES) iDR6 \citep{2022A&A...666A.121R} and show how well we can reconstruct the same stellar parameters {when we fit the model to the photometry and the parallax data.}

We query the GES survey for sources at high Galactic latitude ($|b|>45^\circ$) which have a valid $[Fe/H]$ and $T_{\text{eff}}$ value with a measurement error of less than $0.1$ dex and $110$K for the metallicity and effective temperature, respectively. The GES values for the the stellar parameters estimates and their error are accurate and extinction-independent \citep{2015A&A...577A..77S}, due to the survey being rigorous in deriving the values using multiple methodologies \citep{2022ges}. GES iDR6 has been pre-matched with the Gaia DR3 database, allowing us to query each of our selected targets from the database and retrieve a complete Gaia solution. Furthermore, we cross-match the Gaia sources with 2MASS and ALLWISE as outlined in Section \ref{Data} and make all quality cuts as described therein. Finally, we remove all sources with $\log g < 4.0$ dex {to mitigate against the selection of post-main-sequence stars}, and make an absolute magnitude cut (using the parallax) of $7.5>M_G>4.5$ mag, leaving us with a final sample size of 990 sources. 

We fit our model to the data using an exponential prior on $A_0$ with $1/d_A=5$ $\rm{mag}^{-1}$. We find that there is a systematic offset between our mean effective temperatures and those derived by GES with a median difference in the sample of $136.04$ K. The median value of the standard deviations of the effective temperature across the sample is $183.0$ K. The median difference between the posterior metallicity and the GES-derived metallicity is $0.276$ dex, and The median value of the standard deviations of the posterior metallicity across the sample is $0.21$ dex. The median difference between the log surface gravity derived from the posterior and GES's value is $0.05$ dex, and the median posterior standard deviation across the sample is $0.04$ dex.

Recall that, via the transformation described in Section \ref{transform}, we can transform extinction into effective temperature and leave the point invariant in the observed data space within the errors of the model. We can use this relation to {calibrate our forward model} in our sample assuming that the effective temperatures from GES are complete in accounting for the effective temperatures in an extinction-independent manner. 

\begin{figure*}
    \centering
    \includegraphics[width=\textwidth]{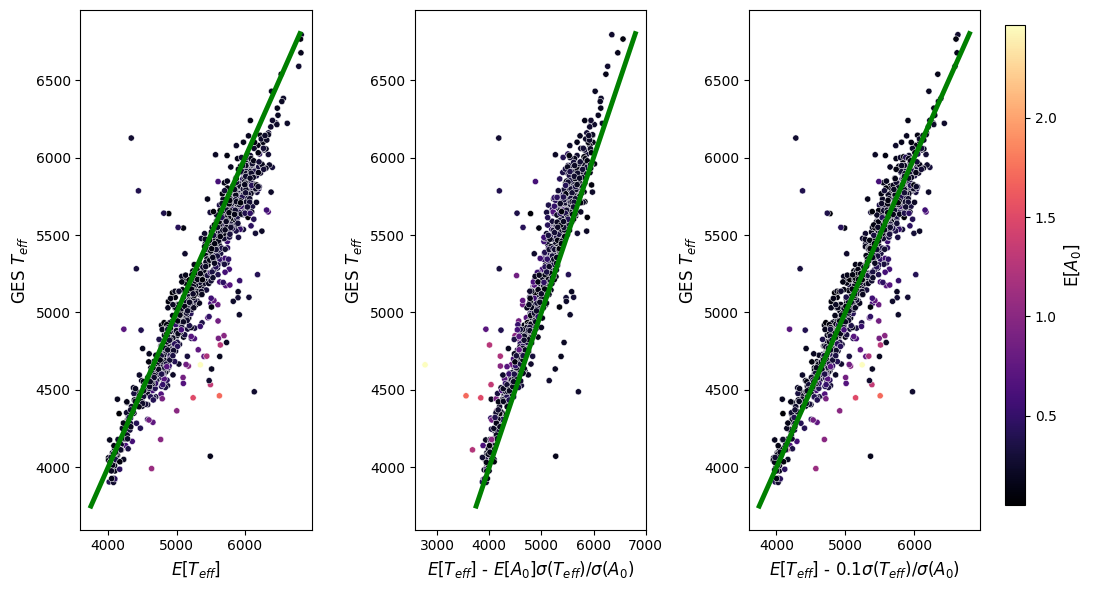}
    \caption{Converting the posterior mean extinctions to the effective temperatures derived by Gaia ESO. From the left, the image shows the effective temperatures (K) derived from GES versus the unaltered posterior mean effective temperature for each point in the sample. The middle plot shows the GES effective temperature versus the mean posterior effective temperature after {adjusting the posterior mean extinction values of each star using the prescription of Section \ref{transform}}. The right plot shows the transformation using the zero-point extinction value which gave the best fit between the transformed effective temperature and the GES effective temperatures. This amounts to subtracting a $\delta A_0=0.1$ mag amount of extinction from the posterior mean effective temperature. The hue indicates the posterior mean extinction (mag) of the star.}
    \label{fig:ges_calib}
\end{figure*}

To illustrate this we show three different cases for our transformed effective temperature in Figure \ref{fig:ges_calib}. From the left, the image shows the effective temperatures derived from GES versus the unaltered posterior mean extinction for each point in the sample. The middle plot shows the GES effective temperature versus the mean posterior effective temperature after {individually {adjusting the posterior mean extinction values of each star using the prescription of Section \ref{transform}}. To generate the third plot, we transformed a fixed amount of $\delta A_0$ to effective temperature for the whole sample and chose} the value which gave the best fit between the transformed effective temperature and the GES effective temperatures. This amounts to taking an $\delta A_0=0.1$ mag amount of extinction, transforming it and subtracting it from the posterior mean effective temperature. {These differences arise because using photometry alone is not sufficient to constrain the extinction-effective temperature degeneracy to the level of GES, due to their use of high-resolution spectroscopy. Thus, our extinction posteriors are wider and posterior mean values of $A_0$ will be systematically larger.}

We can analyse the amount of extinction which remains after this transformation. We find a median value of $A_0=0.12$ mag for this transformed sample. We query the Planck \citep{Planck} dust map for the Galactic coordinates of each of the stars in this sample and find a mean value of $A_0=0.13$ mag. Thus, having highly accurate, independent constraints of effective temperature will be useful in calculating the mean extinction of a set of stars. If we look to regions where we expect negligible extinction and have stars with reliable effective temperatures, we can use this information to infer the distribution of the model inaccuracies in the photometry $\sigma_{\text{floor}}$.

If using spectroscopic estimates of the effective temperature and extinction one needs to be particularly careful what information has been included to derive those values. Extinction values, as point estimates, derived from GES are incompatible with point estimates of the effective temperatures from LAMOST as they assume different assumptions about extinction.

\subsection{Outliers from GES Sample}
\begin{figure}
    \centering
    \includegraphics[width=0.5\textwidth]{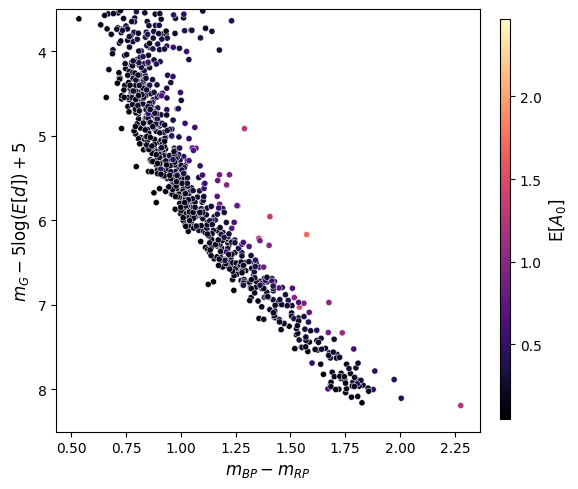}
    \caption{The distance corrected Gaia HR diagram for the region using the Gaia ESO cross-matching. The plot shows a narrow main sequence with a small scatter and a few points, potentially outliers, lying relatively far away from the main sequence. The hue indicates the posterior mean extinction (mag) of the star.}
    \label{fig:hr_outliers_ges}
\end{figure}

We use the same sample as the previous section, derived by cross-matching high Galactic latitude sources from Gaia ESO with Gaia, 2MASS and ALLWISE, to show that high extinction points may be outliers and further analysis should be carried out. We display a plot of the distance corrected Gaia HR diagram for the region in Figure \ref{fig:hr_outliers_ges}. The plot shows a narrow main sequence with a small scatter and a few points with a significant scatter on the right of the main sequence. These points have been characterised as stars with relatively high extinction. We suspect that some of these stars may be binary stars, others young stars and some genuinely high extinction points. 

We stress this because some algorithms choose to smooth spatially over extinction values in the inference step. This can mitigate against outliers, but it can also artificially increase the calculated extinction of a region. Thus, flagging stars with improbable extinctions under a well-chosen prior is necessary for outlier detection. {Through generating binary stars and young stars using our model, these outliers occupy a similar region of the HR diagram as binary or young stars do and we suspect that these outliers are one of these types.}

\section{Concluding Remarks}
{
\subsection{CMB Context}
We conclude this paper with a reminder of the context and a discussion of future works.
Accounting for the effects of non-Gaussian dust distributions for searches of B modes from inflation and a finer structural understanding of Galactic dust is essential for future B mode experiments \citep{2024MNRAS.527.5751A}. The Simons Observatory will map the CMB to 1 arcminute, BICEP operates at 100 GHz and 150 GHz at angular resolutions of $1^\circ$ and $0.7^\circ$ and Planck maps the CMB at resolutions greater than 10 arcminutes. Moreover, the CMB-S4 have a resolution goal of $2$ arcminutes \citep{2022AAS...24021001M}. For each survey's scale of interest, accurate dust distributions along the line-of-sight are of paramount importance for foreground removal. Moreover, line-of-sight polarisation signals which are extrapolated from higher frequency dust maps to lower, CMB-dominated frequencies may be contaminated by 3-dimensional effects which can decorrelate the polarisation maps between different frequencies, reducing the accuracy of empirical extrapolation models below the level required to detect a primordial B-mode signal \citep{2015MNRAS.451L..90T}. There may be small areas of sky (of arbitrary shape) with extremely low dust which could become important calibrators for CMB studies and will be regions of interest for future work.}

\subsection{Star Count at High Galactic Latitudes}

{At high Galactic latitudes $|b|>55^\circ$, we find an average star count within our selection criteria and quality cuts to be $\leq 10$ per arcminute. On the scales of the Simons Observatory, this is too few to recover accurate line-of-sight extinction profiles. However, as we increase the size of the region of interest we can start to find regions with enough stars to construct reliable extinctions along the line-of-sight. If we have reliable knowledge of the Gaia extinction law for the BP/RP spectra, one could drop the cross-matches with longer wavelength passbands when we probe regions of expected low extinction. This is because, in the very low extinction regime cross-matching with longer wavelength passbands will not add significant information if we have a strong prior on the extinction being close to zero. Therefore, the fine variation could be traced by our understanding using only Gaia which would greatly increase the star count, particularly in future releases of Gaia when the data for fainter stars are published.}

\subsection{Stellar Model
Systematics and Calibration}

{To fully utilise Gaia for this small variation, we need to ensure the stellar models are accurately calibrated to the Gaia photometric and spectroscopic data. However, the degeneracy with the effective temperature and other stellar parameters will always be present, so increasing the number of stars we have accurate and reliable effective temperatures at high Galactic latitudes will significantly constrain extinction along the line of sight. Extinction-independent calibration of the Gaia BP/RP spectra should provide more robust joint solutions of effective temperatures, metallicities and extinctions for stars at high Galactic latitude.}{ {In Appendix \ref{panstarrs} we illustrate how narrower passband photometric observations in the Gaia passband range ($\approx  330 \text{nm} - 1050 \text{nm}$) will improve our ability to constrain extinction at high Galactic latitudes. Thus, we expect well-calibrated Gaia BP/RP spectra will significantly improve extinction estimates at high latitudes.}}

{{There are many known systematics which propagate errors to the final extinction posterior. One must account for these errors or the extinction posterior will be unrealistically narrow around a potentially false value. Known systematic errors which arise in the forward modelling process include parallax offsets, intrinsic errors in the chosen stellar model and extinction law, errors in the definition of the passbands, errors arising from interpolating between a grid of values and calibration effects in the spectroscopic estimates of stellar parameters used as data.}}

{When reliably calibrated BP/RP spectra are available to the community this will provide significant information in the optical for each star. In the low-extinction regime, optical observations are sufficient in constraining extinction provided the prior distribution is tight enough to assign a low probability to high-extinction sources due to the NIR photometry becomes less prominent in discerning low extinction variation (Section \ref{sed_fit}).}

\subsection{Extinction Prior}
{{We saw in Section \ref{validation_syn} that extinction is highly degenerate with both the effective temperature and the metallicity, and that constraints on each of these parameters propagate to the extinction posterior distribution. If knowledge about the extinction distribution is known, we can incorporate a relevant prior into our model. At high Galactic latitudes, we expect the extinction to be low, and a natural probability distribution to choose as a prior on the extinction parameter is an exponential distribution. This is a typical choice used to represent knowledge about low extinction sightlines and is common in Type Ia supernovae modelling (see, for example, \cite{2022MNRAS.510.3939M}). High-frequency structures can be included but this is usually derived from dust modelling and a more complex prior is reusing other analysis.}}

{
{For generating large-scale dust maps, it is useful to include the prior information of expecting extinction to be a non-decreasing function of distance along a line-of-sight. \cite{gspphot} use a prior $p(A_0 | d, b) \propto e^{-A_0/\mu}$, where $\mu$ is a function of radial distance and Galactic latitude. Moreover, \cite{greenmap} bin stars into different lines-of-sight and assumes that each extinction bin along the line-of-sight has to be greater than or equal to the previous one. In any case, one has to solve the problem of constraining extinction when it is low. This is because, for any line-of-sight at high Galactic latitudes, there will be at least one star that one needs to constrain the low extinction for and any errors involved in this process will propagate to inferring the extinction of stars at greater distances along that line-of-sight. For single-star inference, this prior must be included as a continuous function of distance. However, many lines-of-sight are dominated by single clouds, where the observable effects do not increase with distance and we have no reason to expect fine variation in the ISM to be a continuous function of distance.}}

\section{Conclusions}

This paper illustrates the complexities which arise when charting low variations in extinction for high Galactic latitude sightlines. We developed a Bayesian model to identify a region of the Hertzsprung-Russell diagram suited to constrain the single-star extinction accurately at high Galactic latitudes. Using photometry from Gaia, 2MASS and ALLWISE together with parallax from Gaia, we employ nested sampling to fit the model to the data and analyse the posterior over stellar parameters for both synthetic and real data. The main conclusions are as follows.

\begin{itemize}

    \item  One has to be careful in choosing priors on stellar parameters. The significant degeneracy means any {prior on the extinction will restrict the posterior effective temperatures}, and vice versa. If prior assumptions are incompatible, or too narrow, one derives a non-physical extinction posterior.
    \item The inaccuracies in the stellar models have a significant impact on constraining the true extinction. Calibration against independent, highly accurate effective temperatures for stars at high Galactic latitudes will constrain the systematic inaccuracies. We propose a cut in Gaia absolute magnitudes, $4.0<M_G<8$, where the stellar models are the most accurate and the systematic effects are minimised.
    \item The extinction-effective temperature degeneracy allows for a family of point estimate pairs of $(T_{\text{eff}},A_0)$, for a single star at high Galactic latitudes, which achieve a similar posterior probability. One can fit a curve to the right plot in Figure \ref{fig:constraint} to convert a pair to another equally likely pair with different extinction.
    \item {The deviation from a tight one-to-one relation between the extinction and effective temperature samples is important for comparing effective temperatures from different surveys.} Effective temperatures can appear to be substantially different. However, this can happen due to different assumptions about the underlying extinction and both solutions can be equally probable under the posterior distribution.
    \item Outliers, such as binary systems or young stars, can appear to have significantly higher extinction than other stars in the sample. If these stars are binned into sightlines to derive extinctions they can cause the extinction of the line-of-sight to be skewed. Flagging these stars is important when we wish to constrain small variations in the ISM.
\end{itemize}

\section*{Acknowledgements}

We are grateful for support from The Gianna Angelopoulos Programme for Science Technology and Innovation (GAPSTI). KSM is supported by the European Union’s Horizon 2020 research and innovation programme under European Research Council Grant Agreement No 101002652 and Marie Skłodowska-Curie Grant Agreement No 873089.

\section*{Data Availability}

This study uses data from Gaia DR3 \citep{GAIADR3}, 2MASS \citep{2MASS} and ALLWISE \citep{ALLWISE}, all of which are publicly available. Moreover, the stellar parameters estimates used throughout come from \cite{Andrae_2023}, LAMOST DR8 \citep{LAMOSTDR8} and Gaia-ESO \citep{2022ges}, which are all publicly available.

\begin{figure*}
    \centering
    \includegraphics[width=\textwidth]{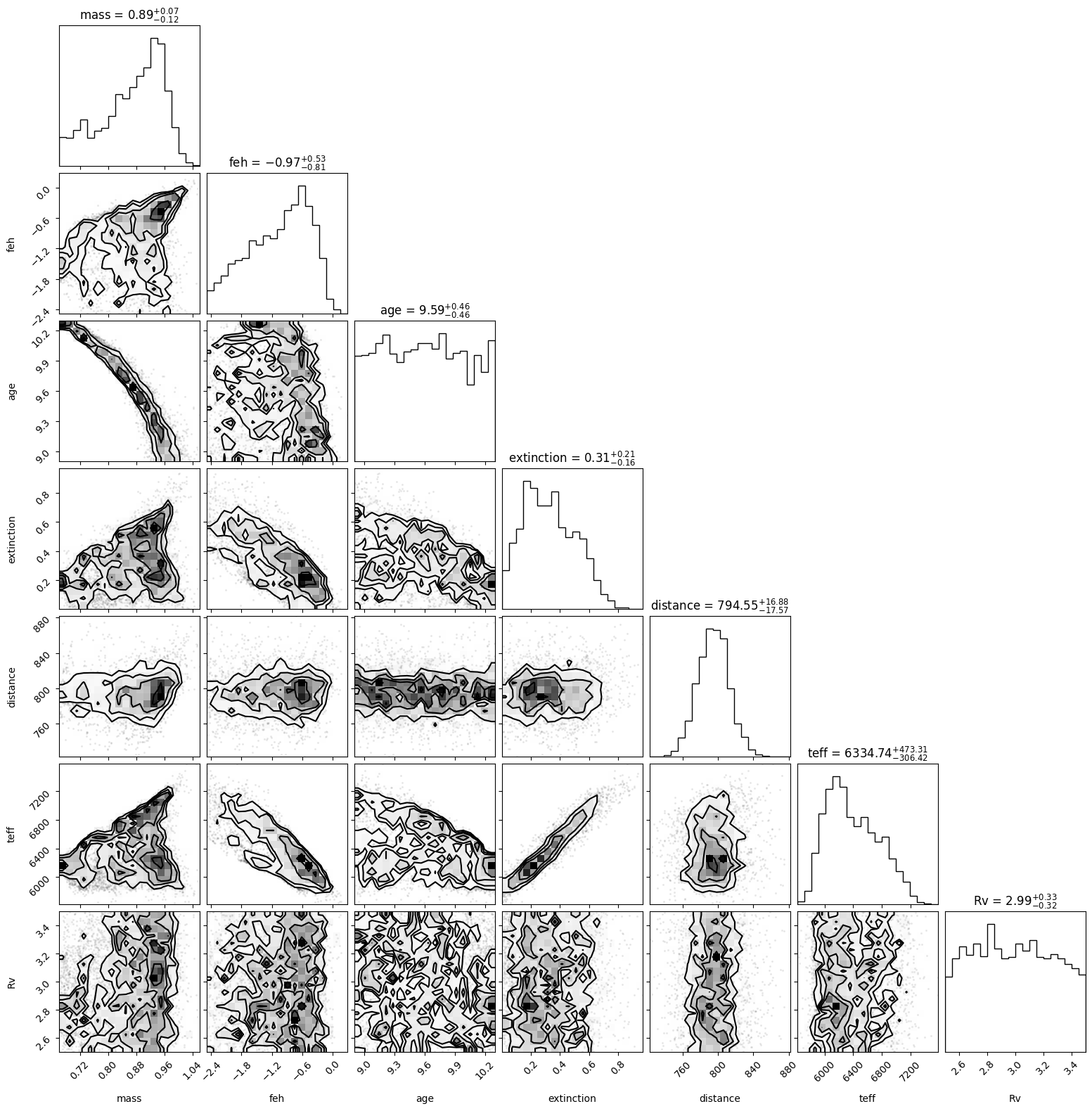}
    \caption{Posterior samples with uniform metallicity prior and no spectroscopic constraints. Corner plot of the posterior distribution for a single star with  $[Fe/H]=-0.125$ dex, $T_{\text{eff}}=5916$ K, age=$10^9$, $M_G=4.7$ mag, $d=800$ pc, $A_0=0.15$ and $m=M_\odot$.}
    \label{fig:corner_uniform}
\end{figure*}

\begin{figure*}
    \centering
    \includegraphics[width=\textwidth]{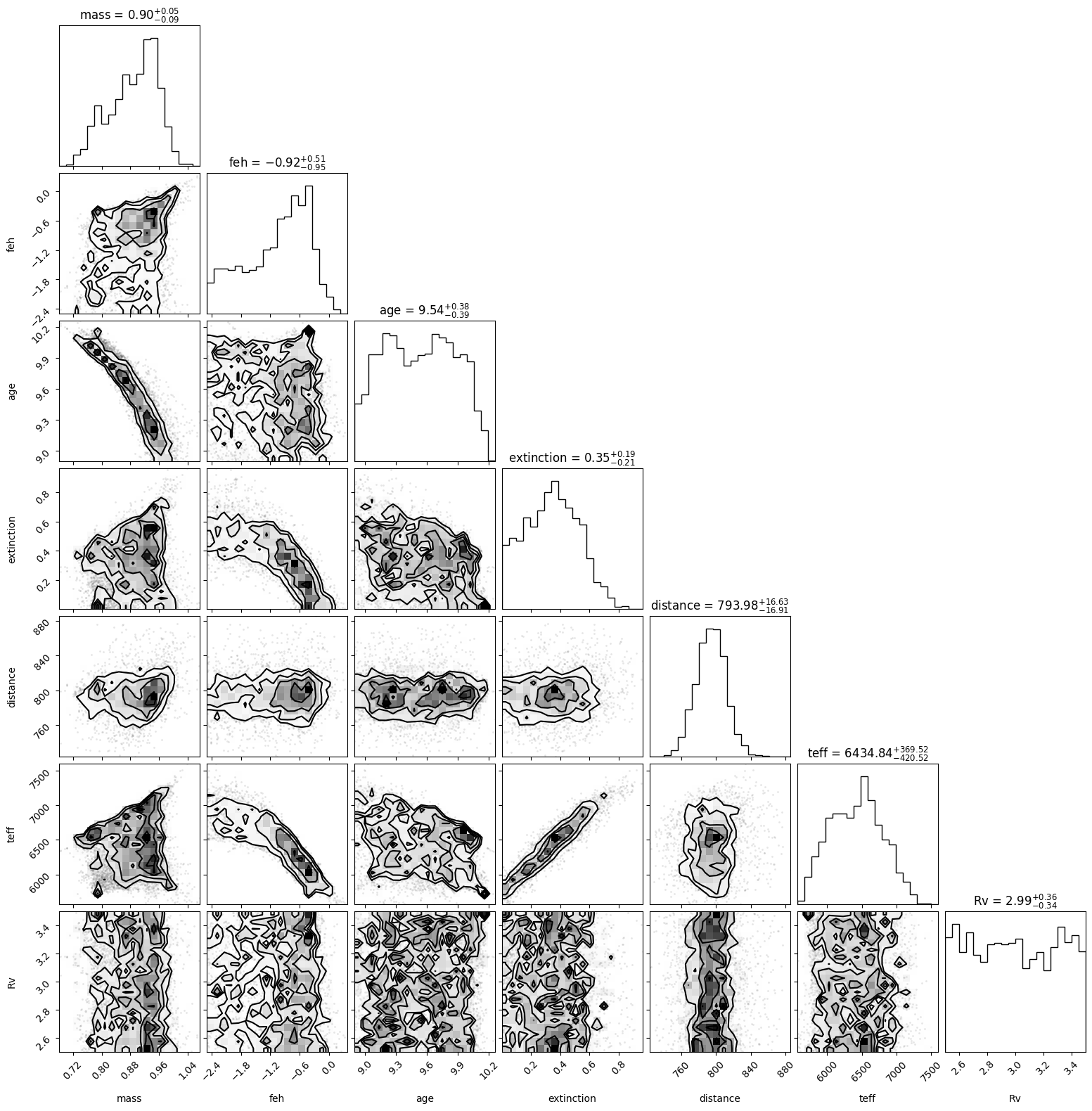}
    \caption{Posterior samples with an astrophysical metallicity prior and no spectroscopic constraints. Corner plot of the posterior distribution for a single star with  $[Fe/H]=-0.125$ dex, $T_{\text{eff}}=5916$ K, age=$10^9$, $M_G=4.7$ mag, $d=800$ pc, $A_0=0.15$ and $m=M_\odot$.}
    \label{fig:corner_uniform_except_metal}
\end{figure*}

\begin{figure*}
    \centering
    \includegraphics[width=\textwidth]{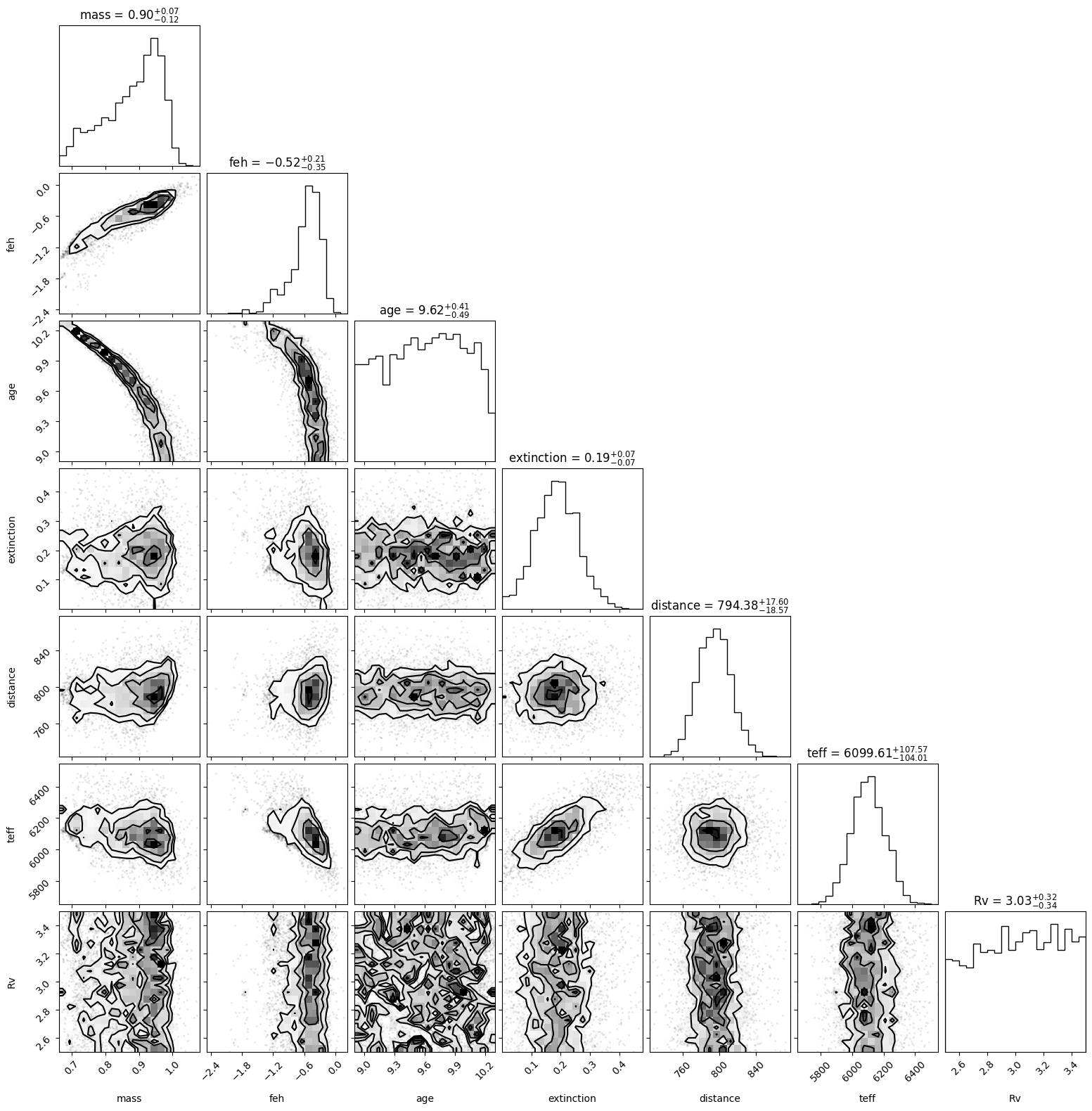}
    \caption{Posterior samples with uniform metallicity prior and a spectroscopic constraint on the effective temperature with $\sigma(T)=110$ K. Corner plot of the posterior distribution for a single star with  $[Fe/H]=-0.125$ dex, $T_{\text{eff}}=5916$ K, age=$10^9$, $M_G=4.7$ mag, $d=800$ pc, $A_0=0.15$ and $m=M_\odot$.}
    \label{fig:corner_uniform_except_metal_teff_cont}
\end{figure*}



\bibliographystyle{mnras}
\bibliography{example} 




\appendix

\section{Using Pan-STARRS Data}\label{panstarrs}
It is well known that the wide passbands in the optical, such as Gaia's, have temperature-dependent extinction coefficients \citep{Zhang_2022}. In this section, we show that under ideal measurement error conditions and assuming no systematic effects are introduced, using narrower passbands in optical wavelength ranges will partially narrow the extinction posterior. 

We illustrate this by performing synthetic analysis using the grizy passband filter from the highly accurate Panoramic Survey Telescope and Rapid Response System 1 (Pan-STARRS, \cite{PS1}) which is a 1.8 m optical and near-infrared telescope located in Hawaii, USA. The mean $5 \sigma$  point source limiting sensitivities in the stacked $3\pi$ Steradian Survey in grizy are (23.3, 23.2, 23.1, 22.3, 21.4) respectively. The upper bound on the systematic uncertainty in the photometric calibration across the sky is 7-12 millimag depending on the bandpass. We synthesise the photometry in the same manner as described in Section \ref{methods} by using the passband definitions as described in \cite{2012ApJ...750...99T}.

To understand how the Pan-STARRS passbands influence the extinction posterior we generate the identical uniform sample as described in Section \ref{uniform}, however, we also include synthetic photometry for the grizy passbands and add extinction in these passbands using the \cite{f99} extinction law with $A_0=0.15$. The error on the photometric magnitudes in the new passbands is synthesised in the same manner as the other passbands as described in Section \ref{uniform}. That is, we model the measurement error using real data from a high Galactic sample.

We fit our model to the data by assuming uniform priors on the metallicity and the extinction parameter (in the same manner as Section \ref{uniform_priors}). Moreover, we assume all photometric magnitudes have the default $\sigma_{\text{floor}}=0.05$ mag. We recall from Table \ref{tab:widths} that without using Pan-STARRS photometry the median difference between the posterior mean extinction and the true value of $A_0=0.15$ is $0.206$. When we fit our model to the data including the Pan-STARRS photometric magnitudes, we find a median difference between the posterior mean and the true value of $0.128$. However, the distance becomes less pronounced when we incorporate metallicity and extinction priors. 

\begin{figure*}
    \centering
    \includegraphics[width=\textwidth]{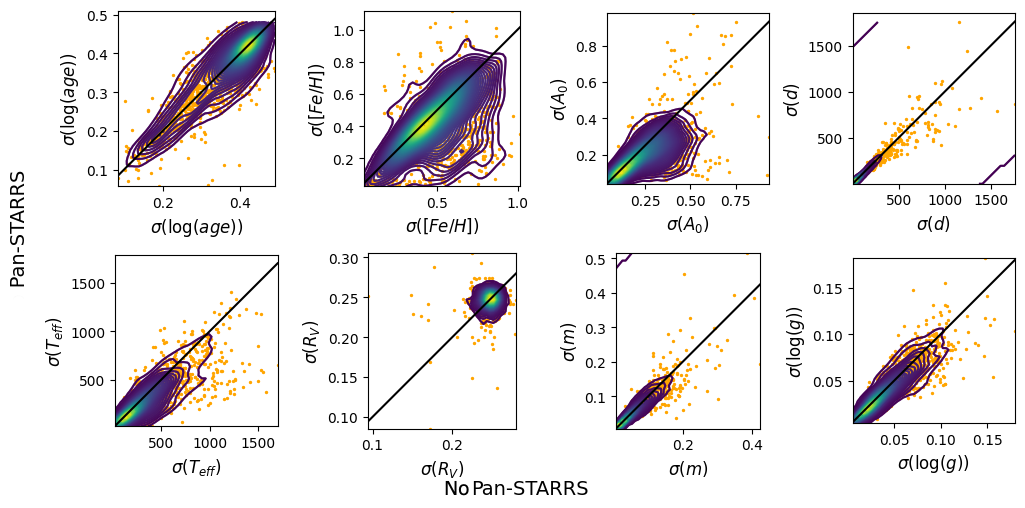}
    \caption{Scatter plot of posterior standard deviation for each parameter over the uniform sample described in Section \ref{uniform} when using uniform priors and (horizontal axis) using the Pan-STARRS data and (vertical axis) not using the Pan-STARRS data.}
    \label{fig:PSvsGaia}
\end{figure*}

We stress that the key results from the paper are independent of using the Pan-STARRS data. However, using any extra photometric passbands that are well calibrated, such as Pan-STARRS, will help us constrain the SED of a star and therefore provide useful information in further constraining the extinction posterior. If we wish to chart very small variations in extinction Pan-STARRS will be useful due to the accuracy of its' photometry and the density of the survey. To conclude we present a plot of the errors derived from the uniform sample both with and without using Pan-STARRS data in Figure \ref{fig:PSvsGaia}

\section{Low Effective Temperature Offset in APOGEE}\label{apogee_appendix}

It has been noted in the literature that the APOGEE \citep{apogeedr16} spectroscopic effective temperature estimates disagree with other surveys, often attributed to an incorrect extinction estimate \citep{2023A&A...670A.107H}. In this section, we show that both LAMOST and APOGEE provide effective temperatures which are consistent with each other, even within the range of disputed effective temperatures, in the sense that we can use the relation in the right plot of Figure \ref{fig:constraint} to transform a certain amount of extinction into an effective temperature correction term to reproduce the LAMOST effective temperatures and the APOGEE effective temperatures.

 We use the sample we generated in Section \ref{crossmatch_LA}, which was generated by cross-matching high Galactic latitude sources from LAMOST with APOGEE, Gaia, 2MASS and ALLWISE. Therein, we showed that we could generate effective temperatures consistent with LAMOST by finding the effective temperatures consistent with a zero extinction solution. We find that taking an extinction contribution of $\delta A_0=0.15$ and finding the effective temperatures corresponding to a global reduction of this extinction provides a good fit between the transformed effective temperatures and those derived by APOGEE. 

\begin{figure*}
    \centering
    \includegraphics[width=\textwidth]{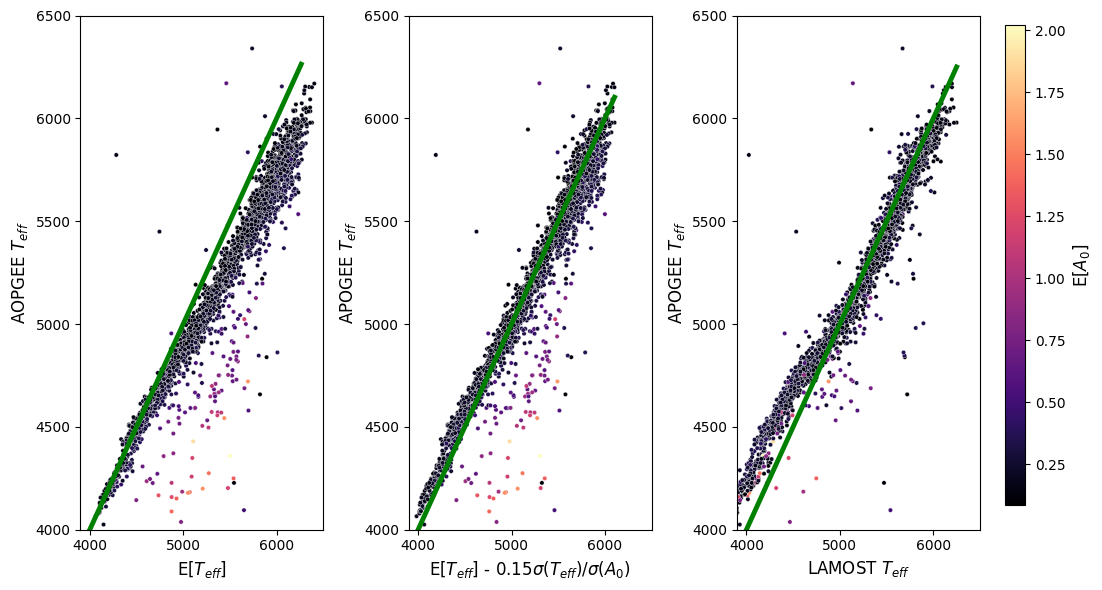}
    \caption{Illustrating APOGEE and LAMOST effective temperatures have the same posterior probability. In Section \ref{crossmatch_LA} we showed that we could extinction-correct our mean effective temperature to derive the LAMOST effective temperatures, both values had similar posterior probabilities. From the left, the image shows the effective temperatures derived from APOGEE versus the unaltered posterior mean effective temperature for each point in the sample. The middle plot shows the transformation using the zero-point extinction, which best fits the transformed effective temperature and the APOGEE effective temperatures. This amounts to subtracting $\delta A_0=0.15$ amount of extinction from the posterior mean effective temperature. The right plot shows the APOGEE effective temperatures versus LAMOST effective temperatures, unaltered by extinction.}
    \label{fig:apogee}
\end{figure*}

Subtracting this contribution, we find a median difference between our transformed mean posterior effective temperatures and their effective temperatures to be $0.17$ K. The mean difference is $28.3$ K. It is important to note that we can reliably transform between the APOGEE and the LAMOST effective temperatures using this relation. The zero point of extinction in both of these cases is different due to prior assumptions, but they both achieve indistinguishable probabilities under the full posterior distribution. We present an illustration of these results in Figure \ref{fig:apogee}.

\bsp	
\label{lastpage}
\end{document}